\pdfoutput=1

\documentclass[12pt,a4paper]{jhepLike}
\let\ifpdf\relax
\usepackage{ifpdf,color}
\let\normalcolor\relax
\usepackage{mathtools,cite}
\usepackage{lipsum}
\usepackage[percent]{overpic}
\usepackage[export]{adjustbox}
\usepackage{array}
\usepackage{amsmath}
\usepackage{graphicx}
\usepackage{bm}
\usepackage{caption}
\usepackage{subcaption}
\usepackage{comment}

\pdfoutput=1 
\usepackage{amsmath,amssymb}

\usepackage{caption}
\usepackage{subcaption}

\DeclareMathOperator{\tr}{Tr}

\newcommand{\eq}[1]{\vspace{-0.5pt}\begin{equation}#1\vspace{-0.5pt}\end{equation}}
\newcommand{\eqst}[1]{\vspace{-0.5pt}\begin{equation*}#1\vspace{-0.5pt}\end{equation*}}
\newcommand{\fwbox}[2]{\text{\makebox[#1][c]{$\hspace{-150pt}\displaystyle#2\hspace{-150pt}$}}}

\newcommand{\fig}[3]{\raisebox{#1}{\ \includegraphics[scale=#2]{#3}}}
\newcommand{\mi}{\raisebox{0.75pt}{\scalebox{0.75}{$\,-\,$}}}
\newcommand{\pl}{\raisebox{0.75pt}{\scalebox{0.75}{$\,+\,$}}}
\renewcommand{\phi}{\varphi}

\newcommand{\x}[2]{x_{#1\hspace{0.1pt}#2}^2}
\renewcommand{\hat}{\widehat}
\definecolor{dim}{rgb}{0.75,0.75,0.75}
\definecolor{hblue}{rgb}{0.18,0.19,0.572}
\definecolor{hred}{rgb}{0.745,0.118,0.176}

\newcommand{\cAA}[2]{\mathcal{A}_{#1;\hspace{0.2pt}#2}}

\newcommand{\z}[2]{\tilde{z}_{#1\hspace{0.5pt}#2}}
\newcommand{\RInv}[5]{[#1\hspace{0.7pt}#2\hspace{0.7pt}#3\hspace{0.7pt}#4\hspace{0.7pt}#5]}

\newcommand{\twoBra}[2]{\langle #1 \hspace{0.5pt} #2 \rangle}

\newcommand{\fourBra}[4]{\langle #1 \hspace{0.5pt} #2 \hspace{0.5pt}#3 \hspace{0.5pt} #4 \rangle}

\newcommand{\fiveBra}[5]{\langle #1 \hspace{0.5pt} #2 \hspace{0.5pt}#3 \hspace{0.5pt} #4 \hspace{0.5pt} #5 \rangle}

\newcommand{\sixBra}[6]{\langle #1 \hspace{0.5pt} #2 \hspace{0.5pt}#3 \hspace{0.5pt} #4 \hspace{0.5pt} #5 \hspace{0.5pt} #6 \rangle}

\newcommand{\sevenBra}[7]{\langle #1 \hspace{0.5pt} #2 \hspace{0.5pt}#3 \hspace{0.5pt} #4 \hspace{0.5pt} #5 \hspace{0.5pt} #6 \hspace{0.5pt} #7 \rangle}

\def\cF{{{\cal F}}}
\def\cN{{{\cal N}}}

\def\cZ{{{\cal Z}}}
\def\cI{{{\cal I}}}

\def\zT{\mathcal{Z}}

\def\Gr{\text{Gr}}
\def\SL{\text{SL}}
\def\GL{\text{GL}}
\def\SU{\text{SU}}

\title{{\Large Multi-Particle Amplitudes from the Four-Point Correlator in Planar $\mathcal{N}=4$ SYM}}
\author{{\normalsize \mbox{Paul~Heslop, Vuong-Viet~Tran}}\\
\mbox{{
\ Centre for Particle Theory, Department of Mathematical Sciences, Durham University,}}\\
\mbox{{
\ South Road, Durham DH1 3LE, United Kingdom}}\\\vspace{-10pt}\\
\mbox{\hspace{-13pt}{{\it E-mails:} {\tt paul.heslop@durham.ac.uk}}}\\\vspace{-15pt}\\\mbox{\hspace{28pt}{ {\tt vuong-viet.tran@durham.ac.uk}}}\vspace{-10pt}}
\date{\today}
\abstract{ A non-trivial consequence of the super-correlator/super-amplitude duality is that the integrand of the four-point correlation function of stress-tensor multiplets in planar $\mathcal{N}=4$ super Yang-Mills contains a certain combination of $n$-point amplitude integrands for any $n$.  This combination is the sum of products of all helicity super-amplitudes with their corresponding helicity conjugates. The four-point correlator itself is described by a single scalar function whose loop level integrands possess a hidden permutation symmetry facilitating its computation up to ten loops. We discover that assuming Yangian symmetry and an appropriate basis of planar dual conformal integrands it is possible to disentangle the contributions from the individual  amplitudes from this combination.
We test this up to  seven points and up to two loops.
This suggests that any scattering amplitude for any $n$, with any helicity structure and at any loop order may be extractable from the four-point correlator. 
}
\preprint{DCPT-18/11}
\newpage
\begin{document}
\vspace{-6pt}\section{Introduction}\label{sec:introduction}\vspace{-4pt}

$\mathcal{N}=4$ Super Yang-Mills theory (SYM) is singled out  amongst all four-dimensional quantum field theories for its remarkable symmetry and mathematical structure along with its key role in  the AdS/CFT correspondence. Further simplification and intriguing structures arise in the planar limit where the number of colours in the gauge group becomes large. 

The fundamental objects of interest in any conformal theory are gauge-invariant operators and their correlation functions. The simplest operator in $\mathcal{N}=4$ SYM is $\mathcal{O}(x)=\tr(\phi(x)^2)$ where $\phi(x)$ is one of the six scalars. This is a special operator: related via superconformal symmetry to both  the stress-energy tensor and the on-shell Lagrangian; protected from renormalisation, and annihilated by half of the supercharges in the theory. Two- and three-point correlators of such operators are protected from renormalisation taking their free value, therefore making the four-point correlator the first non-trivial case. This correlator has been well studied over a number of years and has been computed to three loops fully and to  ten loops at the level of the integrand~\cite{hep-th/9811155,hep-th/9811172,hep-th/9906051,hep-th/0003096,1108.3557,1201.5329,1303.6909,1512.07912,1609.00007}.
It was the discovery of a hidden permutation symmetry at the integrand level---vastly reducing the basis of integrands to a problem of enumerating planar graphs with certain conformal properties~\cite{1108.3557}---which made these results  possible. The basis is given in terms of so-called $f$ graphs, which at $\ell$ loops are graphs with $(4\pl\ell)$-vertices, containing both edges and numerator lines with resulting conformal weight four. 
This provides a compact representation of the correlation function for which we provide examples in subsection \ref{subsec:fgraphs}.
Efficient graphical methods have since been developed for obtaining the coefficients of the $f$ graphs and hence fully determining the correlator.

Another set of key quantities in $\cN=4$ SYM are the integrands of scattering amplitudes, possessing remarkable mathematical structure and at the forefront of current research~\cite{
1303.4734,1505.05886,1704.05460, 1012.6032,1008.2958,ArkaniHamed:book,1512.07912,1609.00007,1312.1163}.
A feature of planar $\mathcal{N}=4$ SYM relates the aforementioned objects to one another via the correlator/amplitude duality. This equates the correlator (divided by its Born contribution) in a polygonal light-like limit to the square of the scattering amplitude normalised by its MHV tree-level value~\cite{1007.3243,1007.3246,1009.2488}. The duality was later extended to incorporate supersymmetry~\cite{1103.3714,1103.4119,1103.4353}.

One can consider this duality for a correlator of any number of points, but it is particularly interesting to apply it to the \textit{four}-point correlator integrand. In this case, taking an $n$-point light-like limit (involving internal integration points as well as ``external'' points---the permutation symmetry means there is no distinction) gives the sum of products of all $n$-point helicity superamplitudes with their helicity conjugates. This remarkable feature makes use of the fact that the $\ell$-loop 4-point correlator integrand is itself an  $n$-point $(\ell{+}4{-}n)$-loop correlator with 4 scalar operators and $n{-}4$ Lagrangians, which are in turn related by supersymmetry to ${\mathcal O}(x)$. 

 Concretely then, taking the $n$-point light-like limit of the $(\ell{+}n{-}4)$-loop, 4-point correlator, represented by  $\mathcal{F}^{(\ell+n-4)}$ (whose precise definition will be given later) we obtain the following combination of N${}^k$MHV, $m$-loop, $n$-point superamplitudes (normalised by the MHV tree-level superamplitude),  $\mathcal{A}_{n;\hspace{0.5pt}k}^{(m)}$,
\eq{\lim_{\substack{\text{n-gon}\\\text{light-like}}} \Big(  \xi^{(n)}\mathcal{F}^{(\ell+n-4)}\Big)
	=\frac{1}{2}\sum_{m=0}^{\ell}\sum_{k=0}^{n-4}\mathcal{A}_{n;\hspace{0.5pt}k}^{(m)}\,\mathcal{A}_{n;\hspace{0.5pt}n-4-k}^{(\ell-m)}/(\mathcal{A}_{n;\hspace{0.5pt}n-4}^{(0)}),
	\label{n_point_dualitya}}
where $\xi^{(n)}$ is a simple algebraic factor.

Note that this sum involves all  N${}^k$MHV amplitudes at $\ell$ loops, as well as lower-loop amplitudes. Furthermore, these are all combined together into a simple scalar function of the external momenta only, $\mathcal{F}^{(\ell)}$, without any complicated helicity/superspace  dependence---the correlator,   $\mathcal{F}^{(\ell)}$, is a {\em much} simpler object than the constituent amplitudes themselves.

{ The question we address in this paper is whether $\cF^{(\ell)}$ contains all the information about these constituent amplitudes, or put another way,  whether one can extract all the individual superamplitudes themselves purely from the combination $\cF^{(\ell)}$.}
We know this can be achieved at four- and five-points~\cite{1108.3557,1201.5329,1312.1163}. This may seem unlikely for higher points at first glance: on the left-hand side, $\cF^{(\ell)}$  is a purely scalar function of the momenta, whereas the superamplitudes on the right-hand side can exhibit complicated helicity structure.

Our findings are consistent with the following conjecture: assuming the tree-level MHV and anti-MHV amplitudes, parity, Yangian symmetry and a dual conformally invariant basis of planar integrands:%
\footnote{At two loops, we use the smaller prescriptive basis of Bourjaily \& Trnka~\cite{1505.05886} for simplicity.} {\em all $n$-point N${}^k$MHV scattering amplitude integrands at any loop order (modulo signs%
	\footnote{The amplitude is fixed up to an overall sign ambiguity for each $0 \!<\!k\!\leq\!(n-4)/2$ which the correlator can never fix. This is because the correlator always gives combinations of the form $\mathcal{A}_{n;\hspace{0.5pt}k}\mathcal{A}_{n;\hspace{0.5pt}n{-}k{-}4}$ which is invariant under the simultaneous transformations: $\mathcal{A}_{n;\hspace{0.5pt}k} \rightarrow - \mathcal{A}_{n;\hspace{0.5pt}k},\  \mathcal{A}_{n;\hspace{0.5pt}n{-}k{-}4} \rightarrow - \mathcal{A}_{n;\hspace{0.5pt}n{-}k{-}4}$. However, we stress that this ambiguity is an overall sign for the entire all-loop amplitude that can be fixed once and for all at tree level. There is then also 
		an additional overall sign ambiguity for the entire parity-odd sector of the MHV/anti-MHV amplitude for a similar reason. This second type of sign ambiguity can be fixed once and for all at 1 loop.\label{footnote2}})
	 can be obtained from the four-point correlator, which thus packages all of this information together into a simple scalar function.}

Let us now make the above statement more precise and specify what information can be obtained from the correlator at each loop level.
First note that  the $(\ell{+}n{-}4)$-loop correlator 
combination~\eqref{n_point_dualitya} involves the parity-even $\ell$-loop combinations (consider $m=0$ and $m=\ell$ in~\eqref{n_point_dualitya})
\eq{{{\mathcal A}_{n;\hspace{0.5pt}k}^{(\ell)}{\mathcal A}_{n;\hspace{0.5pt}n-k-4}^{(0)} + {\mathcal A}_{n;\hspace{0.5pt}n-k-4}^{(\ell)}{\mathcal A}_{n;\hspace{0.5pt}k}^{(0)} }={{\mathcal A}_{n;\hspace{0.5pt}k}^{(\ell)}\overline{\mathcal A}_{n;\hspace{0.5pt}k}^{(0)} + \overline{\mathcal A}_{n;\hspace{0.5pt}k}^{(\ell)}{\mathcal A}_{n;\hspace{0.5pt}k}^{(0)},}
\label{parityeven}}
together with lower-loop amplitudes.
Thus from this combination alone, the correlator at this loop level cannot see ambiguities in the amplitude of the form
\eq{{\mathcal A}_k^{(\ell)} \rightarrow {\mathcal A}_k^{(\ell)} + {\mathcal A}_k^{(0)} \cI_{\text{k-ambiguity}}^{(\ell)} , \qquad \qquad 
		{\mathcal A}_{n-k-4}^{(\ell)} \rightarrow {\mathcal A}_{n-k-4}^{(\ell)} - {\mathcal A}_{n-k-4}^{(0)} \cI_{\text{k-ambiguity}}^{(\ell)},
\label{amb}}
where $\cI_{{\text{k-ambiguity}}}^{(\ell)}$ is any combination of ${\ell}$-loop integrands.\footnote{Note that for the special case of $k\!=\!n\!-\!k\!-\!4$, this ambiguity is absent. This is the case for NMHV six points as we shall see later. \label{footnote3}}
Remarkably, we find that~\eqref{amb} is the {\em only} form of ambiguity arising from the duality at this loop level, and even  more remarkably, this ambiguity is resolved by considering the correlator at one loop higher.
Imposing parity reduces the ambiguity $\cI_{\text{k-ambiguity}}^{(\ell)}$ to the space of parity-odd integrands only, and imposing cyclicity further reduces this to just the space of cyclic combinations of parity-odd integrands.

More precisely then, the conjecture is that knowing all $n$-point amplitudes fully to  $(\ell{-}2)$ loops, as well as  the parity-even pieces (as defined in~\eqref{parityeven}) at $(\ell{-}1)$ loops, then from the light-like limit of the $(\ell{+}n{-}4)$-loop four-point correlator, we can extract the ``parity-even'' part of all $\ell$-loop amplitudes along with fixing the remaining ambiguities at $(\ell{-}1)$ loops. Thus, we can recursively extract the parity-even part of the $\ell$-loop amplitude from the $m$-loop correlator, $\cF^{(m)}$ with $m=1, \ldots, \ell\pl n \mi 4$, and the entire amplitude if we additionally use $\cF^{(\ell+n-3)}$.

In this paper, we verify this statement by checking at six points  and seven points up to two loops for the parity-even part.

In order to achieve this, we use  a basis of planar dual conformal $\ell$-loop integrands, $\cI_j^{(\ell)}$ to construct an ansatz for the superamplitudes. 
The integrands are functions of $\x{a}{b}$, where $x_a,x_b	$ are dual momenta (external or loop co-ordinates) together with the parity-odd dual conformal covariant, most straightforwardly expressed as $\varepsilon(X_{a_1},X_{a_2},X_{a_3},X_{a_4},X_{a_5},X_{a_6})$ where the $X_a$ are 6d embedding dual momentum co-ordinates.
At two loops we use a refinement of this basis, namely 
the prescriptive basis of  \cite{1303.4734,1505.05886,1704.05460} which we show can be written in terms of the above basis.

We also need to control the helicity  structures of the superamplitudes. For this we use a basis of Yangian-invariant Grassmannian integrals, $R_{k;\hspace{0.5pt}i}$ and as a technical aid, we use amplituhedron co-ordinates~\cite{0909.0250,0912.3249,1312.2007,1312.7878}.
We thus write an ansatz for the constituent superamplitudes of the form
\eq{\mathcal{A}_{n;\hspace{0.5pt}k}^{(\ell)} = \sum_{ij}\alpha_{ij}  R_{k;\hspace{0.5pt}i} \hspace{0.5pt} \cI^{(\ell)}_j  ,\label{ansatz}}
which we substitute into the duality equation~\eqref{n_point_dualitya} in order to determine the coefficients $\alpha_{ij}$.

To extract the combination~\eqref{n_point_dualitya} from this ansatz, we use a number of tools: dual-space, momentum twistors, the Grassmannian, amplituhedron co-ordinates, rules from the amplituhedron, and an understanding of the duality in question. For completeness, we review these ideas---starting with $f$ graphs and the correlator/amplitude duality. We then review momentum twistors, the Grassmannian, amplituhedron co-ordinates and proceed with the six-point extractions. 

Finally, we repeat the idea for seven particles which invokes the Grassmannian to derive invariants and draw conclusions from our results.
%
\vspace{-6pt}\section{Review of Key Ideas}\label{sec:review}\vspace{-6pt}
%
\vspace{-2pt}\subsection{Representing the Correlator  with $f$ graphs}\label{subsec:fgraphs}\vspace{-0pt}
The four-point correlator of interest is
\eq{\mathcal{G}_4(x_1,x_2,x_3,x_4)\equiv\langle\mathcal{O}(x_1)\overline{\mathcal{O}}(x_2)\mathcal{O}(x_3)\overline{\mathcal{O}}(x_4)\rangle,\label{correlator_definition}}
involving the protected operator $\mathcal{O}(x) = \tr(\phi^2)$ for any scalar $\phi$ in the theory. We define $\mathcal{F}^{(\ell)}$ as the the $\ell$-loop \textit{integrand} of  (\ref{correlator_definition}) 
\eq{\mathcal{G}^{(\ell)}_4(x_1,x_2,x_3,x_4;x_5, \dots ,x_{4+\ell})= 2 x_{1\hspace{0.1pt}3}^4x_{2\hspace{0.1pt}4}^4 \mathcal{F}^{(\ell)}(x_1, \dots x_{4+\ell}),}
where $x_{5}, \ldots, x_{4+\ell}$ are the $\ell$-loop integration variables. Note that we have a unique definition of this integrand via the $(4{+}\ell)$-point correlator involving $\ell$ Lagrangian insertions in addition to the four operators in~\eqref{correlator_definition}.

There is a powerful hidden symmetry~\cite{1108.3557} possessed by this integrand (for $\ell>0$) which simply states that $\cF^{(\ell)}$ is invariant under any  $S_{4+\ell}$ permutation of its variables. This therefore places ``external'' variables $x_1, x_2, x_3, x_4$ on the same level as internal variables, $x_5, \ldots,  x_{4+\ell}$. The symmetry vastly reduces the basis of potential integrands for the correlator to that of objects $f_i^{(\ell)}$,
\eq{\mathcal{F}^{(\ell)}\equiv\sum_{i}c^{\ell}_i\,f^{(\ell)}_i.\label{f_graph_expansion}\vspace{-5pt}}
These basis elements $f_i^{(\ell)}$ are equivalent to so-called $f$ graphs,  which at $\ell$ loops are \textit{undirected} graphs with $(4\pl\ell)$-vertices composed of both solid (denominator), dashed (numerator) lines and signed degree (number of edges minus number of numerator lines leaving each vertex) equal to four. The solid edges must contribute to a simple planar graph. These graphs and their coefficients have been determined up to ten loops using combinations of various efficient graphical methods~\cite{1609.00007}.
The above provides a compact representation of the correlation function with expressions up to four loops displayed below
\eq{\begin{array}{rc@{$\;\;\;\;\;$}rc@{$\;\;\;\;\;$}rc}\\[-32pt]f^{(1)}\equiv&\fwbox{85pt}{\hspace{10pt}\fig{-42.5pt}{0.35}{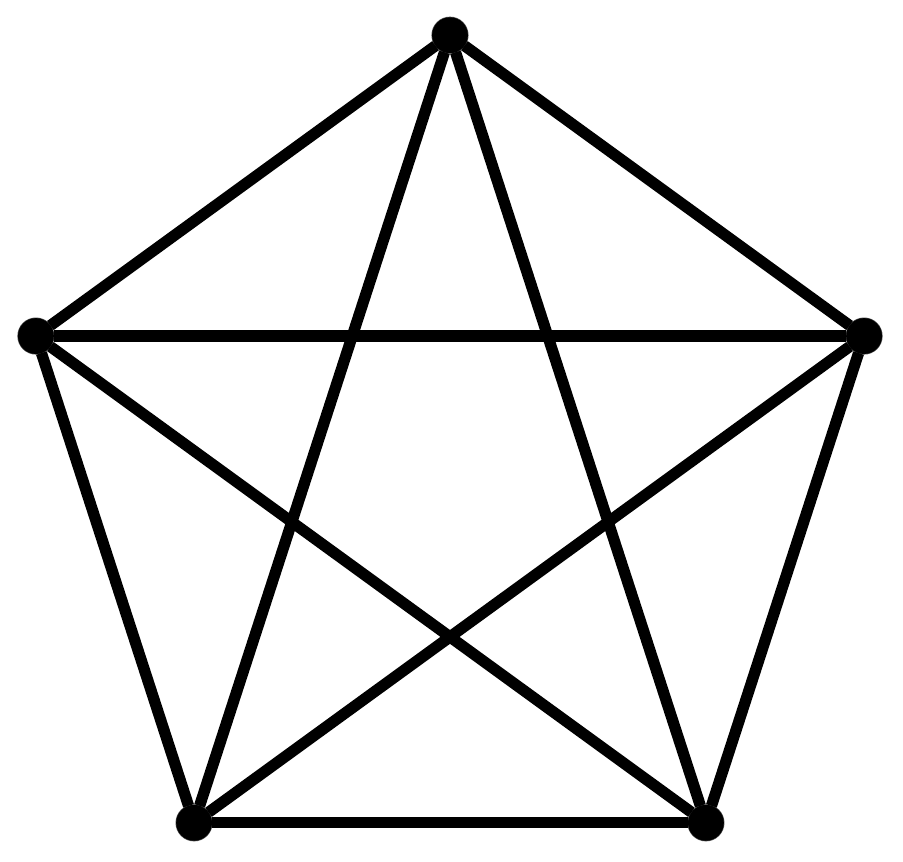}}&f^{(2)}\equiv&\fwbox{85pt}{\hspace{-28pt}\fig{-44pt}{0.35}{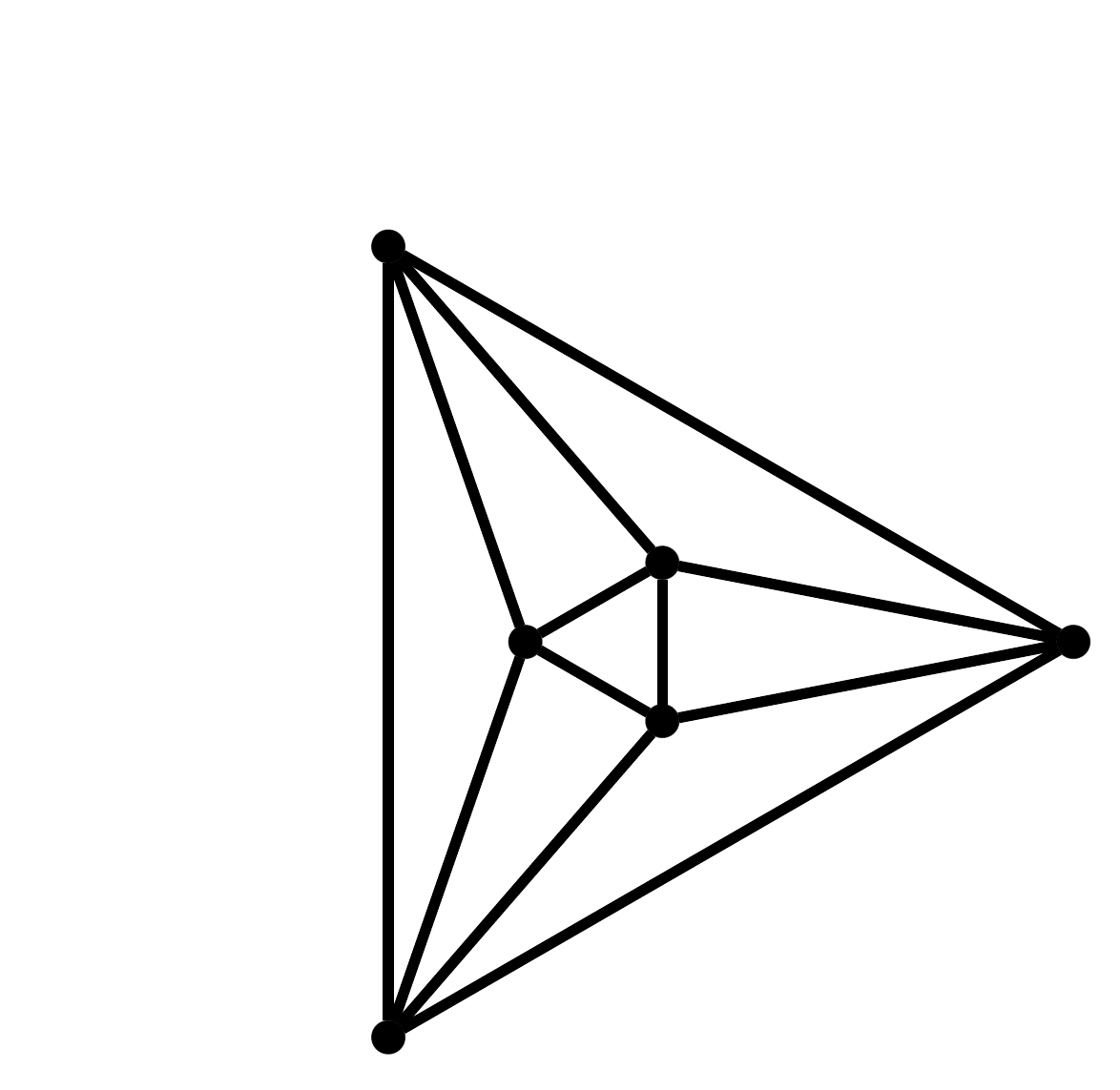}}&f^{(3)}\equiv&\fwbox{85pt}{\hspace{-28pt}\fig{-44pt}{0.35}{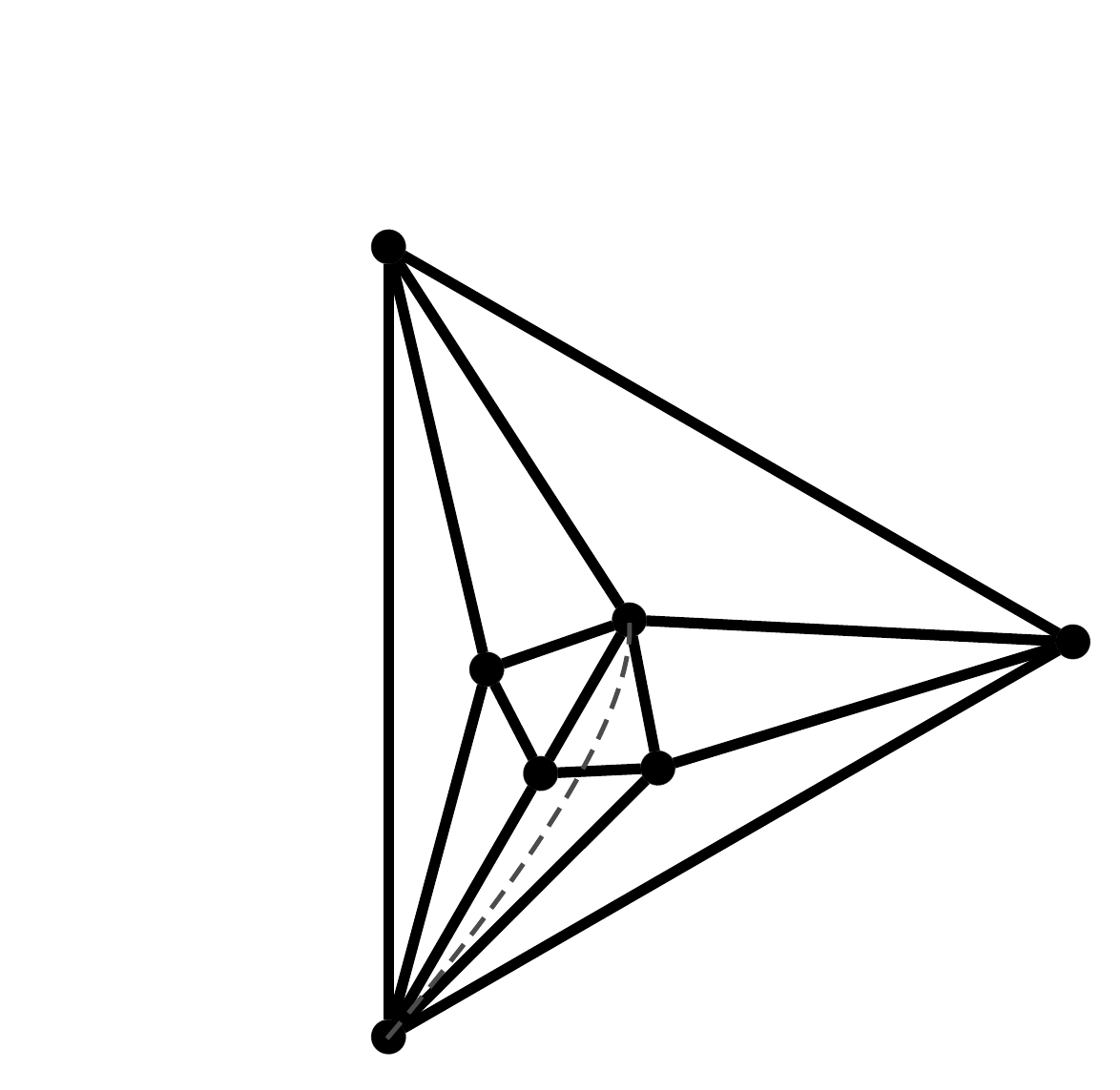}}\\[38pt]f^{(4)}_1\equiv&\fwbox{85pt}{\hspace{-28pt}\fig{-44pt}{0.35}{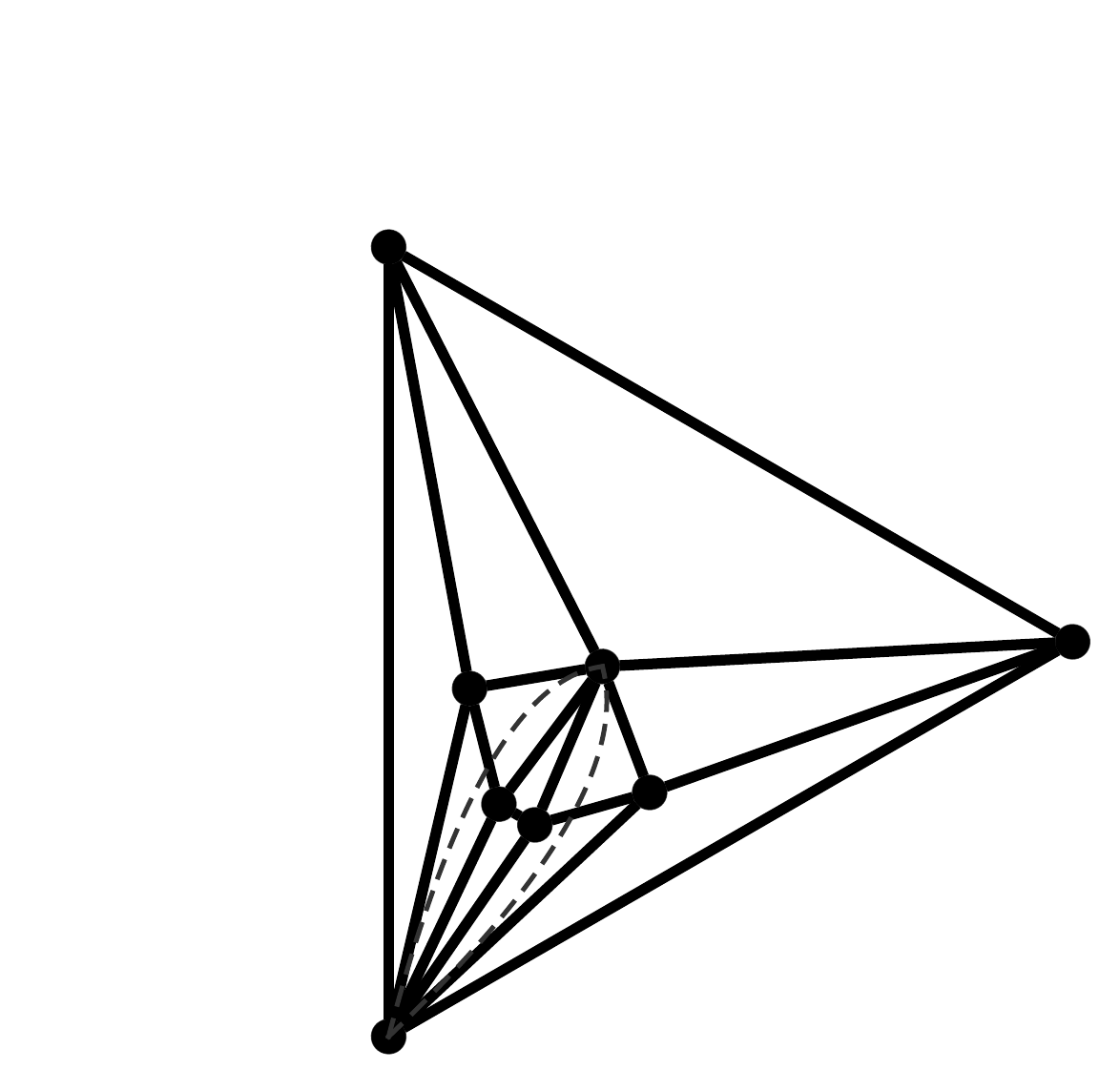}}&f^{(4)}_2\equiv&\fwbox{85pt}{\hspace{-28pt}\fig{-66pt}{0.35}{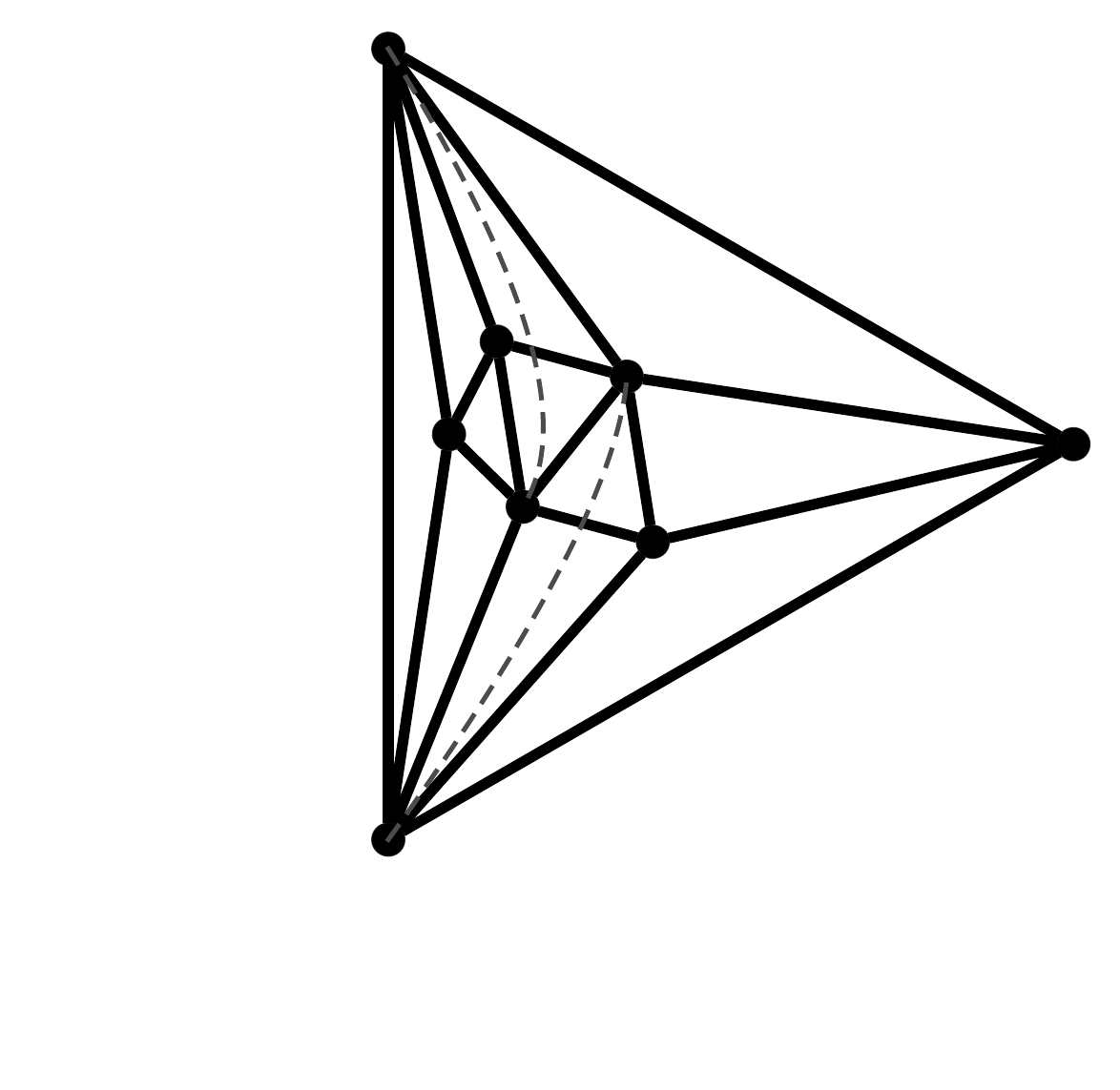}}&f^{(4)}_3\equiv&\fwbox{85pt}{\hspace{14pt}\fig{-42.5pt}{0.35}{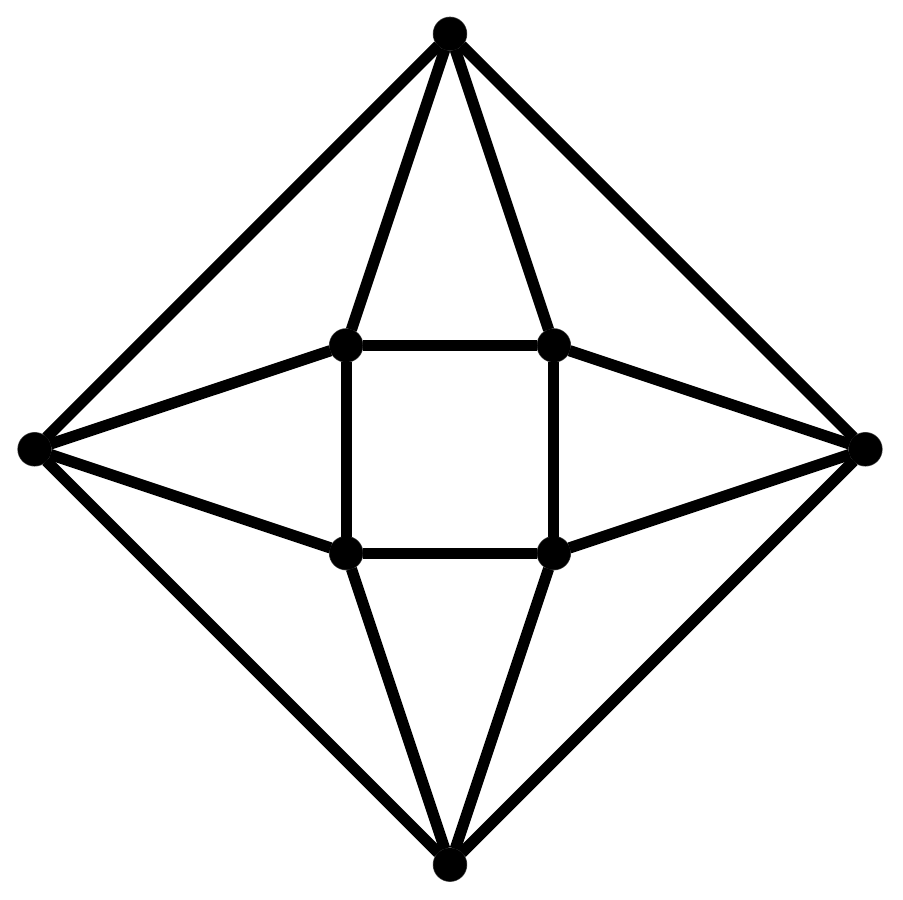}}\\[-35pt]\end{array}\vspace{40pt}\label{one_through_four_loop_f_graphs}\vspace{-0.2in}}
where $\mathcal{F}^{(1)}=f^{(1)}$, $\mathcal{F}^{(2)}=f^{(2)}$, $\mathcal{F}^{(3)}=f^{(3)}$ and $\mathcal{F}^{(4)}=f^{(4)}_1 + f^{(4)}_2 - f^{(4)}_3$ all come with unit-magnitude coefficients up to four loops. Each graph corresponds to an algebraic expression  (we are not careful to distinguish the graphs and their corresponding expressions) as follows:
label the vertices with labels 1 to $4{+}\ell$. Then a solid or dashed edge between vertices $a$ and $b$ is represented by $1/\x{a}{b}$ or $\x{a}{b}$, respectively.  The property of undirectedness is apparent by the property $\x{a}{b}=\x{b}{a}$. Then sum over all possible inequivalent labellings of the graph.

For example, the three-loop correlator is explicitly given as
\vspace{-0.33in}\eqst{\hspace{-80pt}\fig{-65pt}{0.5}{fGraphThreeLoops}\hspace{-1pt}= \frac{{1 \over 20}  \left( (\x{1}{2})^2 \x{3}{4} \x{3}{7} \x{4}{5} \x{5}{6}\x{6}{7}
  + S_7 \text{ permutations}\right)}{\prod_{a<b}\x{a}{b}}\,.\hspace{-50pt}\vspace{0.035in}} 
The above example for $f^{(3)}$ has a single
dashed line coming from the $(\x{1}{2})^2$ in the
numerator which is only partially cancelled by the denominator.

Summing over inequivalent labellings is the same as summing over all permutations and dividing by the size of the automorphism group (symmetry factor) of the graph (20 in the above example). This representation keeps graph labels implicit, so that a given unlabelled graph is defined by a labelled expression summed over all symmetric permutations.

\vspace{-2pt}\subsection{The Correlator/Amplitude Duality}\label{subsec:CorrelatorAmplitude}\vspace{0pt}
The fully supersymmetric correlator/amplitude duality is conjectured to relate integrands for the $m$-point super-correlator to the square of $m$-point super-amplitude integrands in an $m$-gon light-like limit. However, the maximally nilpotent piece of the $m$-point super-correlator is equivalent to the 4-point correlator. We thus consider  the   four-point correlator, $\mathcal{F}^{(\ell)}$  under various $n$-gon light-like limits for $n \geq 4$. Remarkably, this correlator contains information about \textit{all} higher point amplitudes in the $n$-gon limits \cite{1103.3714,1103.4353,1312.1163}. 

It is natural to study amplitude integrands as functions of external and loop momenta. It was however observed in \cite{hep-th/0607160} that it is often useful to reparametrise to ``dual momenta'' via $p_a = x_{a+1} - x_a$ where $x_a$ are understood mod $n$. 
 These dual momenta $x_a$ of the amplitude are identified with the Minkowski co-ordinates $x_a$ of the correlator via the duality. The conformal invariance of correlation functions then implies a hidden dual conformal invariance known to be a property of planar amplitudes in $\cN=4$ SYM~\cite{0707.0243,0807.1095,0807.4097}.

The simplest case of the duality then states that the $4$-point light-like limit (where four consecutive points become light-like separated, $\x{1}{2},\x{2}{3},\x{3}{4},\x{1}{4} \rightarrow 0$) outputs four-particle amplitudes squared. At the integrand level at fixed loop level, we get
\eq{\lim_{\substack{\text{4-gon}\\\text{light-like}}} \Big( \xi^{(4)}  \mathcal{F}^{(\ell)} \Big)=\frac{1}{2}  \sum_{m=0}^{\ell}\mathcal{A}_4^{(m)}\mathcal{A}_4^{(\ell-m)}
		,\label{4_point_duality}}
for $\xi^{(4)}=\x{1}{2}\x{2}{3}\x{3}{4}\x{1}{4}(\x{1}{3}\x{2}{4})^2$. Note that both sides are interpreted as integrands: on the right-hand side,  $\mathcal{A}^{(m)}$ depends on $m$ integration variables and $\mathcal{A}^{(\ell-m)}$ depends on the remaining $(\ell-m)$ integration variables. The result is then symmetrised over all $\ell$  integration variables. Furthermore, all amplitudes (throughout the paper) are understood to be divided by the corresponding tree-level MHV superamplitude.

At higher points, it is convenient to consider superamplitudes written in the  chiral superspace formalism of dual Minkowski superspace~\cite{0807.1095}: 
\eq{\lambda_a^{\alpha} \eta_a^I \equiv \theta_{a+1}^{\alpha I} -\theta_a^{\alpha I},  \label{super_momentum_chiral_superspace}}
for $\SU(4)$ index $I\!=\!1,2,3,4$, where $\eta_a^I$ are  Nair Grassmann-odd super-momentum variables and $\lambda_a^{\alpha}$ spinor helicity variables. In complete analogy to region momenta, we define ``dual/region super momenta'', $\theta_a^{\alpha I}$ so that super-momentum conservation is trivialised.

The higher-point generalisation of~\eqref{4_point_duality} then involves the $n$-point light-like limit ($\x{1}{2},\x{2}{3},$ $\dots , \x{n}{1} \rightarrow 0$) giving 
\eq{\lim_{\substack{\text{n-gon}\\\text{light-like}}} \Big(  \xi^{(n)}\mathcal{F}^{(\ell+n-4)}\Big)
 =\frac{1}{2}\sum_{m=0}^{\ell}\sum_{k=0}^{n-4} \frac{\mathcal{A}_{n;\hspace{0.5pt}k}^{(m)}\,\mathcal{A}_{n;\hspace{0.5pt}n-4-k}^{(\ell-m)}}{\mathcal{A}_{n;\hspace{0.5pt}n-4}^{(0)}}.\label{n_point_duality}} 
Here ${\mathcal A}_{n;\hspace{0.5pt}k}^{(\ell)}$ is the superamplitude divided by the tree-level MHV superamplitude  so that $ \mathcal{A}_{n;\hspace{0.2pt}0}^{\text{tree}}=1$, and $\xi^{(n)}$ generalises $\xi^{(4)}$ as follows
\eq{\xi^{(n)}\equiv\prod_{a=1}^n\x{a}{a+1}\x{a}{a+2}.\label{xi_general_definition}}
Similarly to the four-point case, on  the right-hand side of \eqref{n_point_duality}, $ {\mathcal A}_{n;\hspace{0.5pt}k}^{(m)}$ is understood to depend on $n$ external variables $x_a,\theta_a$ together with $m$ loop variables $x_a$ and $ {\mathcal A}_{n;\hspace{0.5pt}n-4-k}^{(\ell{-}m)}$ to depend on the same external variables, but the other  $(\ell{-}m)$ loop variables---these are then symmetrised over.
Note that the numerator on the right-hand side is a maximally nilpotent superconformal invariant. Since there is a unique maximally nilpotent invariant, this is proportional to the maximally nilpotent  invariant amplitude  $\mathcal{A}_{n;\hspace{0.5pt}n-4}^{(0)}$ and therefore the ratio in~\eqref{n_point_duality} makes sense and removes all $\theta$ dependence. We will see explicit examples of the use of this equation shortly and find amplituhedron variables to be the most useful way of dealing with the Grassmann-odd structure.  

\vspace{-2pt}\subsection{(Super) Momentum Twistors and Amplituhedron Coordinates}\label{subsec:twistors_invariants}\vspace{0pt}

It is convenient to recast in terms of   
 ``momentum twistors''~\cite{0909.0250, 0905.1473}.
Complex Minkowski space is equivalent to the Grassmannian of 2-planes in $\mathbb{C}^4$,  $X_{\alpha}^{A}\in \Gr(2,4)$ where $\alpha=1,2$, $A=1,\ldots,4$ and
\eq{
	X_{\alpha}^A \sim M_\alpha{}^\beta X_{\beta}^A,}
for any $\GL(2)$ matrix $M$.
Here the rows $X_{1}^A$ and $X_2^A$ are basis elements of the 2-plane and the equivalence relation corresponds to a change of basis. One choice of basis fixes the first 2x2 block to the identity and the next 2x2 block corresponds to standard co-ordinates for Minkowski space (in spinor notation) 
\eq{
X_{\alpha}^B \sim (\delta_\alpha ^\beta,  x_{\alpha \dot \beta}).}
Two Minkowski co-ordinates that are light-like separated correspond to two planes which intersect. In the case of the light-like limit of the correlator where we have $n$ consecutively light-like separated co-ordinates, it makes sense to choose the basis for the corresponding 2-planes to be the lines of intersection. Thus we have 
\eq{
	X_{a \alpha}^A \sim \left( \begin{array}{l}z_{a-1}^A\\z_a^A\end{array}\right),}
where $a=1,\ldots,n$ is the particle number. In this case, the $z_a^A$ are known as momentum twistors.
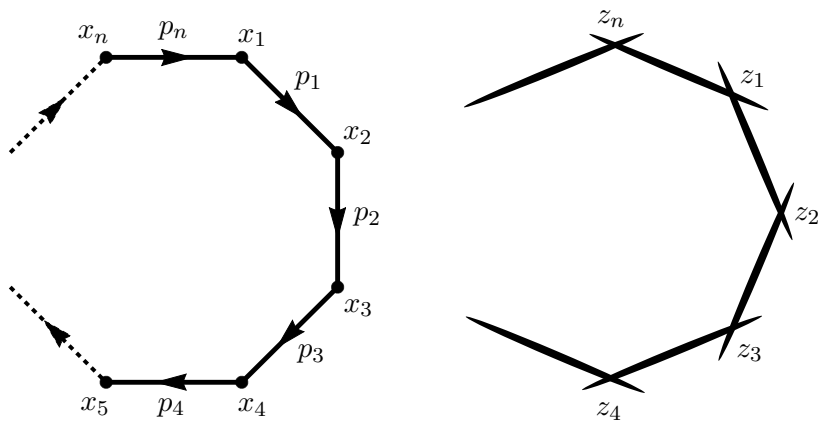
\begin{figure}[h!] \vspace{-0.4in}
  \centering
  \begin{minipage}[b]{0.4\textwidth}
    \eqst{\begin{overpic}[width = 6.5cm]{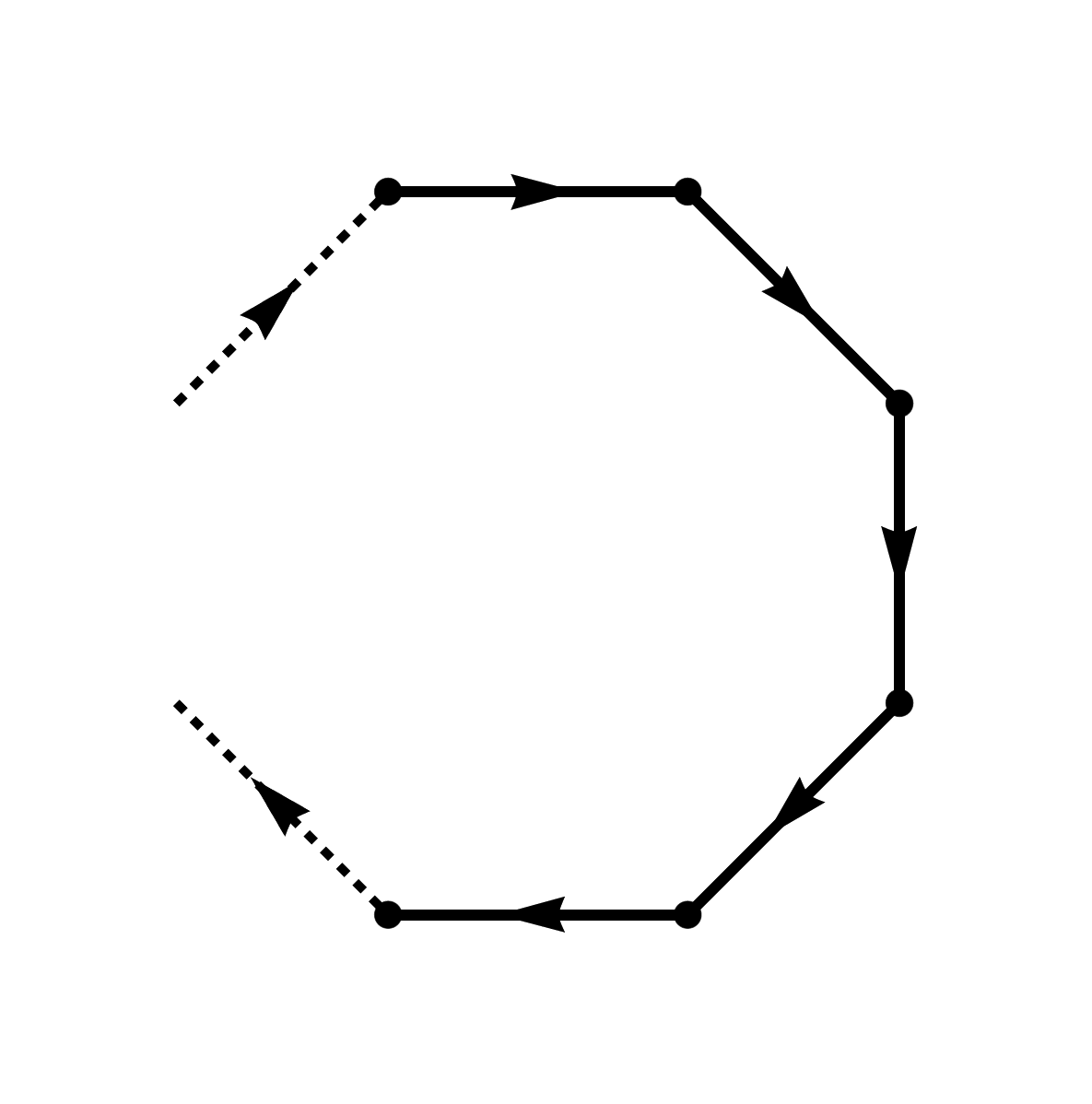}
 \put (29.5,86.5) {\small $x_n$}
 \put (61.5,85.8) {\small $x_1$}
 \put (82.6,66.1) {\small $x_2$}
  \put (82.7,31.3) {\small $x_3$}
 \put (61.5,11.3) {\small $x_4$}
 \put (30,11.5) {\small $x_5$}

 \put (45.7,87.4) {\small $p_n$}
 \put (73,77.2) {\small $p_1$}
 \put (84.7,49.8) {\small $p_2$} 
 \put (73.5,22.2) {\small $p_3$} 
 \put (45.5,11.4) {\small $p_4$}  
 \end{overpic}}
  \end{minipage}
    \hspace{-1cm}
  \begin{minipage}[b]{0.4\textwidth}
    \eqst{\begin{overpic}[width = 6.6cm]{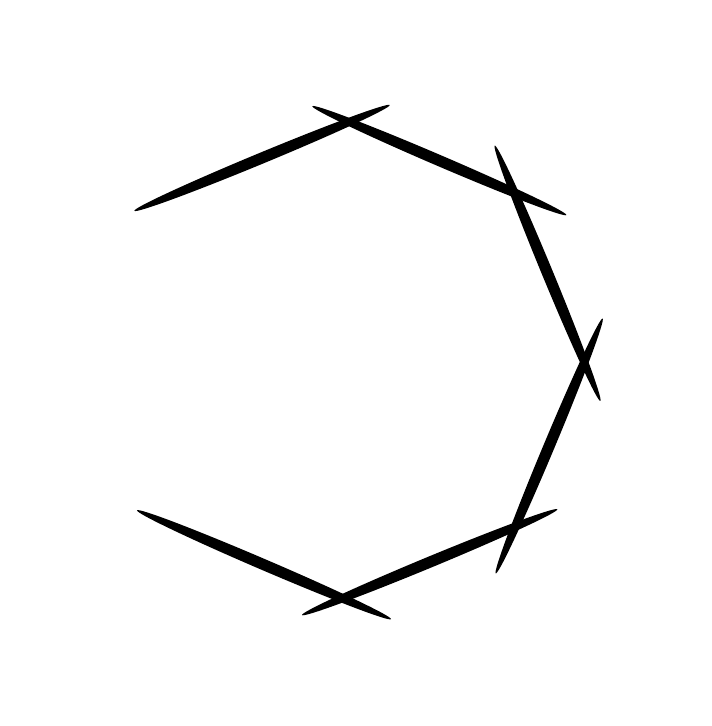}
   \put (44.5,87.55) {\small $z_n$}
  \put (72.5,75.8) {\small $z_1$}
 \put (83.1,48.8) {\small $z_2$} 
 \put (71.8,21.8) {\small $z_3$} 
 \put (44.3,10.2) {\small $z_4$} \end{overpic}}
  \end{minipage} 
  \captionsetup{width=.8\linewidth} \vspace{-0.1in}
  \caption{The transformation between dual-momentum space and momentum-twistor space.} \vspace{-0.1in} \label{dual_momentum_to_twistor} 
  \end{figure}

One often considers projective twistor space $\mathbb{P}^3$ and so points in dual momentum space are associated to projective lines in $\mathbb{P}^3$ which intersect if the corresponding space-time points are light-like separated.
 Thus for the $n$-gon light-like limit we get the picture  illustrated in Figure~\ref{dual_momentum_to_twistor}.
Loop variables in $x$ space also correspond to lines in momentum-twistor space which will not intersect with other lines. These can be 
specified by two twistors each in the same way. 

Dual conformal symmetry acts linearly on the $A$ index of the momentum twistors and so a natural dual conformal covariant is provided by the (momentum) twistor four-bracket defined as the determinant of the square matrix formed by four twistors
\eq{
\fourBra{a}{b}{c}{d}\equiv\text{det}\{z_a, z_b , z_c, z_d\} \propto \varepsilon_{A B C D}\,z_a^A  z_b^B z_c^C  z_d^D. \label{twistor_four_bracket_definition}}
Another way of thinking of Minkowski  space is to take an $\SL(2)$ invariant combination of the $\Gr(2,4)$ co-ordinates 
\eq{\label{6dX}
	X^{AB}=X_{\alpha}^AX_{\beta}^B\varepsilon^{\alpha\beta}.}
Since $X^{AB}$ is anti-symmetric in its indices, it is equivalent to a 6-vector and these are precisely the 6d embedding co-ordinates for Minkowski space.
Notice that the above relations imply 
\eq{\varepsilon_{ABCD}X_a^{AB}X_b^{CD}  \sim \fourBra{a{-}1}{\,a}{\,b{-}1}{\,b} \sim \x{a}{b}.\label{xij2totwbr}
}
In a similar way, chiral superspace can be thought of as the Grassmannian of 2-planes in $\mathbb{C}^{4|4}$
\eq{
	X_\alpha{}^{\mathcal{A}} \sim M_\alpha{}^\beta X_\beta{}^{\mathcal{A}},}
and the entire discussion above gets similarly uplifted into $\mathbb{C}^{(4|4)}$.
We obtain 
super-momentum twistors, $\zT_a^{\mathcal{A}}$, living in $\mathbb{C}^{4|4}$,
\eq{\zT^{\mathcal{A}}_a 
\equiv \big(z_a^A; \chi_a^I \big)\in \mathbb{C}^{4|4}. \label{super_momentum_twistor_definition}}
Beyond the MHV sector, dual superconformal symmetry implies that the superamplitudes can be written in terms of dual superconformal invariants~\cite{0807.1095}. For example, at the NMHV level, these are known as $R$ invariants and defined by a (dual) conformal ratio of four brackets and a Grassmann-odd delta function 
\eq{\RInv{a}{b}{c}{d}{e} \equiv \frac{\delta^{4}\big(\chi_a \fourBra{b}{c}{d}{e} +\chi_b \fourBra{c}{d}{e}{a} + \chi_c \fourBra{d}{e}{a}{b} +\chi_d \fourBra{e}{a}{b}{c} +\chi_e \fourBra{a}{b}{c}{d}\big)}{\fourBra{a}{b}{c}{d} \fourBra{b}{c}{d}{e} \fourBra{c}{d}{e}{a} \fourBra{d}{e}{a}{b}\fourBra{e}{a}{b}{c} }.  \label{R_invariant_delta_definition}}
It is also convenient to further change variables to ``bosonised dual momentum superspace co-ordinates'' or ``amplituhedron co-ordinates'' following~\cite{0905.1473,1312.2007}. For an N${}^k$MHV superamplitude, these variables live in a $(4+k)$-dimensional purely bosonic space and are defined as
\eq{\label{exttw}	
	\hat Z_a\equiv(z_a,\,\chi_a.\phi) \in \mathbb{C}^{4+k},}
where we have introduced the global Grassmann-odd variables $\phi^\alpha_I$, $\alpha=1,\ldots,k$ which converts  the $\chi$ co-ordinate into additional $k$  augmented momentum-twistor co-ordinates. 

For example, for $k=1$, we convert supermomenta to 5-dimensional co-ordinates, 
then
 the superconformal invariant~\eqref{R_invariant_delta_definition} becomes the ratio of five-brackets and four-brackets:
\eq{\RInv{a}{b}{c}{d}{e}\equiv \frac{\fiveBra{a}{b}{c}{d}{e}^4}{\fourBra{a}{b}{c}{d} \fourBra{b}{c}{d}{e} \fourBra{c}{d}{e}{a} \fourBra{d}{e}{a}{b}\fourBra{e}{a}{b}{c}}.
\label{R_invariant_definition}}
Here the 5-brackets are the obvious 5d generalisations of the 4-brackets:
\eq{\fiveBra{a}{b}{c}{d}{e}\equiv\text{det}\{ \hat Z_a,  \hat Z_b,  \hat Z_c ,  \hat Z_d, \hat Z_e\}. \label{Z_bracket}}  
This rewriting will be useful later as it will trivialise  the multiplication of $R$ invariants which we will need when considering products of amplitudes. The expression~\eqref{R_invariant_definition}  becomes equal to~\eqref{R_invariant_delta_definition}  after integrating out the additional $\phi$ co-ordinate.

Note that the anti-MHV $n$-point tree-level superamplitude has the following  simple form in amplituhedron co-ordinates
\eq{{\mathcal A}_{n;\hspace{0.5pt}n-4}^{(0)} = 	\frac{\langle 1\hspace{0.5pt} 2 \dots n\rangle^4}{\fourBra{1}{2}{3}{4}\fourBra{2}{3}{4}{5}\dots \fourBra{n}{1}{2}{3}}.\label{antiMHV}}
%

\vspace{-2pt}\subsection{Yangian Invariants from the Grassmannian} \label{subsec:Yangian_invariants_from_Grass}
\vspace{0pt}
We will need to expand higher $k$ amplitudes in terms of higher $k$ analogues of the $R$ invariants~\eqref{R_invariant_delta_definition}, \eqref{R_invariant_definition}. For any $k$, these superconformal (indeed Yangian) invariants can be understood as residues of a  Grassmannian integral in planar $\mathcal{N}\!=\!4$ SYM \cite{0912.3249,0912.4912,1002.4625,1002.4622,1212.5605}. The main goal here is to introduce the tools needed to take the residues of the Grassmannian, directly in  amplituhedron space and thus derive \textit{covariant} forms for higher $k$ analogues of the $R$ invariants~\eqref{R_invariant_definition}.  Let us therefore introduce the Grassmannian representation for $n$-particle N$^k$MHV Yangian  invariants \cite{1212.5605}: 
\eq{
	\frac{1}{\text{vol}\hspace{0.8pt}[\GL(k)]} \int \frac{d^{\hspace{1pt}k\times n} C_{\alpha a}}{(1 \cdots k)(2 \cdots k\hspace{-1.5pt}+\hspace{-1.5pt}1)\cdots (n \cdots k\hspace{-1.5pt}-\hspace{-1.5pt}1)}\prod_{\alpha=1}^k\delta^{4|4}(C_{\alpha a} \cZ_{a}^{\mathcal{A}}). \label{grassmannian_integral}}
$C_{\alpha a}$ is the $k\!\times\!n$ matrix defining a Grassmannian of $k$-planes in $n$ dimensions, $\Gr(k,n)$ and  $\cZ_{a}^{\mathcal{A}}$ are super twistor  co-ordinates.
The $\GL(k)$-redundancy reflects a change of basis for $k$ planes. The denominator is simply given as $k$-minors constructed from columns of $C$
\eq{(a_1 \cdots a_k )= \det\{C_{\alpha 1},\ldots,C_{\alpha k}\}. \label{grassmannian_minors}}
We need a contour of integration. Note that the integral is $k \!\times \!(n{-}k)$ dimensional (after division by vol$[\GL(k)]$), and there are $4k$ bosonic delta functions, leaving  $k\! \times \!(n{-}k{-}4)$ non-trivial integrals. The non-trivial contributions to these integrals arise from  $k \!\times \!(n{-}k{-}4)$-dimensional poles of the integrand. A spanning set of all possible integrals of this form is thus provided by the residues of these poles, which define a codimension  $k\! \times\! (n{-}k{-}4)$ integration region. This then corresponds to a $4k$ dimensional ``cell'' of $\Gr(k,n)$. These are
in turn classified by permutations (see~\cite{1212.5605}, in particular section 12). From this formalism, one can obtain (positive)\footnote{Positive means the ordered minors of the Grassmannian matrix are all strictly positive if and only if $\alpha_i>0$.} canonical co-ordinates $\alpha_1, \ldots, \alpha_{4k}$ for this cell inside $\Gr(k,n)$ such that the measure in~\eqref{grassmannian_integral}, 
reduces to
the simple $4k$ dlog form
\eq{ \Omega_{k(n-k)} \equiv \frac{1}{\text{vol}\hspace{0.8pt}[\GL(k)]} \frac{d^{\hspace{1pt}k\times n} C_{\alpha a}}{(1 \cdots k)(2 \cdots k\hspace{-1.5pt}+\hspace{-1.5pt}1)\cdots (n \cdots k\hspace{-1.5pt}-\hspace{-1.5pt}1)} \quad  \longrightarrow  \quad \Omega_{4k}=\frac{\text{d}\alpha_1 \dots \text{d}\alpha_{4k}}{\alpha_1 \dots \alpha_{4k}}.\label{dlog}}
Now  we wish to write these Yangian invariants in amplituhedron co-ordinates (which in particular makes multiplying invariants together much simpler). In amplituhedron co-ordinates, the Grassmannian integral~\eqref{grassmannian_integral}, translates simply to 
\eq{\int \Omega \, \delta^{4k}(Y;Y_0).\label{yang}}
Here we have defined 
\eq{Y_{\alpha}^{\tilde{A}} \equiv C_{\alpha a}\hat Z_{a}^{\tilde{A}} , \qquad Y_{0\alpha}^{\tilde{B}} \equiv \left( 0_{\alpha}^B , \delta_{\alpha}^{\beta}\right),   \label{Y_constraints_general}} 
where $\hat Z$ is defined in~\eqref{exttw} and we have split the $4\hspace{-1.5pt}+\hspace{-1.5pt}k$ index $ {\tilde B}$ into an ordinary twistor index and $k$ additional indices $\tilde B=(B,\beta)$. Note that $Y \in \Gr(k,k{+}4)$, and $\delta^{4k}(Y;Y_0)$ is the natural Grassmannian invariant $\delta$-function whose precise definition can be found in~\cite{1312.2007}.

The natural brackets in amplituhedron space, $\mathbb{C}^{4+k}$, are $(4{+}k)$-brackets,  but using   $Y \in \Gr(k,k+4)$ we can form $(4{+}k)$-brackets with four $\hat Z_a$'s and $Y$, for example
\eq{\fiveBra{Y}{a}{b}{c}{d}\equiv  \fiveBra{Y_1 \cdots Y_k}{\hat Z_a}{\hat Z_b}{\hat Z_c}{\hat Z_d} \equiv \text{det}\{Y_1, \ldots, Y_k,  \hat Z_a,  \hat Z_b,  \hat Z_c, \hat Z_d\}. \label{Y_bracket}} 
We could equally replace $Y$ in (\ref{Y_bracket}) with $k$ $\hat Z$s to form $Y$-independent $(4{+}k)$-brackets.

There is an efficient way to arrive at a fully covariant form for a Yangian invariant corresponding to a particular residue via the canonical co-ordinates for this residue. To do this, we think of the reduced measure $\Omega_{4k}$ as a differential form on $Y \in \Gr(k,k+4)$ (simply a change of co-ordinates).
Therefore
\eq{\Omega_{4k} = \twoBra{Y}{d^{\hspace{0.5pt}4} Y_1}\cdots \twoBra{Y}{d^{\hspace{0.5pt}4} Y_k} \times \mathcal{Y}_{n;\hspace{0.5pt}k}(\hat Z_1,\dots,\hat Z_n, Y), \label{Y_weightless_form}} 
where $\mathcal{Y}_{n;\hspace{0.5pt}k}$ is a function of weight $-(k\pl4)$ in $Y$, rendering $\Omega_{4k}$  $Y$-weightless. 
Here $\twoBra{Y}{d^{\hspace{0.5pt}4} Y_1}\cdots \twoBra{Y}{d^{\hspace{0.5pt}4} Y_k}$ is the natural Grassmannian invariant measure, using~\eqref{Y_bracket} but with the anti-symmetric differential form $d^4Y_i$ in the last 4 slots of the $(4{+}k)$-bracket.  

If we can write $\Omega_{4k}$ in this way, the Yangian invariant~\eqref{yang} is simply
\eq{\int \Omega_{4k} \, \delta^{4k}(Y;Y_0) = \mathcal{Y}_{n;\hspace{0.5pt}k}(\hat Z_1,\dots,\hat Z_n, Y_0),}
noting that the brackets involving $Y$ then reduce to 4-brackets  $\fiveBra{Y_0}{a}{b}{c}{d}=\fourBra{a}{b}{c}{d}$.

In fact, we will be able to jump directly from the canonical co-ordinates and corresponding dlog form~\eqref{dlog} to the Yangian invariant $\mathcal{Y}_{n;\hspace{0.5pt}k}(\hat Z_1,\dots,\hat Z_n, Y)$ by a covariantisation procedure. We illustrate this with the example of seven-point $k=2$ Yangian invariants  in \mbox{section \ref{sec:extracting_seven_point_integrands}}.

Note that the amplituhedron (bosonised) form for super-invariants have a number of advantages over the standard form. In particular, non-trivial identities which are very hard to see in the superspace formalism arise naturally as Schouten-like identities of the bosonised quantities. One potential question  is how to extract components from this form. There is a straightforward way to think of this without first converting back to the standard form for the super-invariant in terms of $\chi$'s.
This is particularly straightforward if we seek a component of the form $\chi_a^4\chi_b^4$. Such components are extractable in a canonical way by placing the points $a,b$  adjacent to one another in the six-bracket representation and simply removing them, thus projecting to four brackets, e.g.
\eq{\begin{aligned}\big(\sixBra{1}{2}{3}{4}{5}{6}\fourBra{1}{2}{3}{7}\hspace{-2.5pt}-\hspace{-2.5pt}\sixBra{1}{2}{3}{4}{5}{7}\fourBra{1}{2}{3}{6})^4\big\vert_{\chi_1^4\chi_3^4}&=(-\sixBra{1}{3}{2}{4}{5}{6}\fourBra{1}{2}{3}{7}+\sixBra{1}{3}{2}{4}{5}{7}\fourBra{1}{2}{3}{6})^4 \big\vert_{\chi_1^4\chi_3^4} \\ &= (-\fourBra{2}{4}{5}{6}\fourBra{1}{2}{3}{7}+\fourBra{2}{4}{5}{7}\fourBra{1}{2}{3}{6})^4. \vspace{-0.5pt}\end{aligned}\notag}
%
%

\vspace{-6pt}\section{Six-Point Integrands}\label{sec:extracting_six_point_integrands}\vspace{-4pt}
Let us now consider  the hexagonal light-like limit of the four-point correlator, taking six  points of the correlator to be consecutively light-like separated: $\x{1}{2}\!=\!\x{2}{3}\!=\!\x{3}{4}\!=\!\x{4}{5}\!=\!\x{5}{6}\!=\!\x{1}{6}\!=\!0$. The duality, (\ref{n_point_duality}) becomes
\eq{\lim_{\substack{\text{6-gon}\\\text{light-like}}} \Big(  \xi^{(6)}\mathcal{F}^{(\ell+2)}\Big)
	=\sum_{m=0}^{\ell} \frac{\mathcal{A}_{6;\hspace{0.2pt}0}^{(m)}\,\mathcal{A}_{6;\hspace{0.2pt}2}^{(\ell-m)}    +\frac{1}{2}\mathcal{A}_{6;\hspace{0.2pt}1}^{(m)}  \mathcal{A}_{6;\hspace{0.2pt}1}^{(\ell-m)}   }{\mathcal{A}_{6;\hspace{0.2pt}2}^{(0)}}.\label{six_point_lightlike_correlator}}
We will  restrict this statement to various orders of perturbation, using the known correlator to predict amplitude integrands on the right-hand side. This leads to a simple linear algebra problem for matching coefficients from a sensible ansatz for the amplitude to the known correlator. 

\vspace{-2pt}\subsection{Tree Level}\label{subsec:extracting_tree_level_integrand}\vspace{-0pt}
At tree level, $\ell=0$, the duality (\ref{six_point_lightlike_correlator}) becomes
\eq{\lim_{\substack{\text{6-gon}\\\text{light-like}}}\!\! \xi^{(6)}\mathcal{F}^{(2)}=
	1+\frac{1}{2}\left(\cAA{6}{1}^{(0)}\right)^2 / \cAA{6}{2}^{(0)},\label{six_point_lightlike_correlator_two_loops}}
recalling that all amplitudes are understood to be divided by the tree-level MHV amplitude and thus $\cAA{6}{0}^{(0)}=1$. Evaluating the left-hand side of (\ref{six_point_lightlike_correlator_two_loops}) amounts to symmetrising $f^{(2)}$ over $S_6$, multiplying by $\xi^{(6)}= \x{1}{2}\x{2}{3}\x{3}{4}\x{4}{5}\x{5}{6}\x{6}{1}\x{1}{3}\x{2}{4}\x{3}{5}\x{4}{6}\x{5}{1}\x{6}{2}$ and applying the 6-gon limit. Using (\ref{one_through_four_loop_f_graphs}), one straightforwardly obtains
\eq{\lim_{\substack{\text{6-gon}\\\text{light-like}}}\!\! \xi^{(6)}\mathcal{F}^{(2)}=1+\frac{\x{1}{5}\x{2}{4}}{\x{1}{4}\x{2}{5}}+\frac{\x{2}{6}\x{3}{5}}{\x{2}{5}\x{3}{6}}+\frac{\x{1}{3}\x{4}{6}}{\x{1}{4}\x{3}{6}}.\label{six_point_lightlike_correlator_two_loops_x}}
Equating  (\ref{six_point_lightlike_correlator_two_loops}) and (\ref{six_point_lightlike_correlator_two_loops_x}) then gives  a prediction for  $\frac{1}{2}\big(\cAA{6}{1}^{(0)}\big)^2 / \cAA{6}{2}^{(0)}$ in terms of finite cross ratios. We now wish to derive the NMHV tree-level amplitude itself, $\cAA{6}{1}^{(0)}$ from this combination. We start with an ansatz for  $\cAA{6}{1}^{(0)}$ in terms of $R$ invariants (the six-point Yangian invariants).
At six points, an $R$ invariant in the square-bracket notation is uniquely specified by the index it is missing 
\eq{R_a \equiv \RInv{b}{c}{d}{e}{f} = \frac{\fiveBra{b}{c}{d}{e}{f}^4}{\fourBra{b}{c}{d}{e}\fourBra{c}{d}{e}{f}\fourBra{d}{e}{f}{b}\fourBra{e}{f}{b}{c}\fourBra{f}{b}{c}{d}},}
we will use this notation for the rest of this section.
These six $R$ invariants are not independent since 
\eq{R_1-R_2+R_3-R_4+R_5=R_6,\label{Rid}}
	 so we use only five of these in our basis. 
Thus we have the following ansatz
\eq{\cAA{6}{1}^{(0)}=\alpha_1R_1+\alpha_2R_2+\alpha_3R_3+\alpha_4R_4+\alpha_5R_5, \label{NMHV_tree_ansatz}}
with arbitrary coefficients $\alpha_a$.
Since $R_a^2=0$, the square is:
\eq{\left(\cAA{6}{1}^{(0)}\right)^2=2\sum_{a<b}\alpha_a \alpha_b R_a R_b.  \label{NMHV_tree_ansatz_squared}}
To proceed, we need a rule for multiplying two NMHV $R$ invariants to produce the numerator of the unique six-point N${}^2$MHV invariant. In the numerator of the above, we have combinations such as 
\eq{\fiveBra{a}{b}{c}{d}{e}^4 \fiveBra{a}{b}{c}{d}{f}^4= \sixBra{a}{b}{c}{d}{e}{f}^4 \fourBra{a}{b}{c}{d}^4. \label{amplituhedron_rule_one}}
For six external points, there is a unique non-trivial six bracket. As the above is N${}^2$MHV, the right-hand side must contain $\sixBra{1}{2}{3}{4}{5}{6}^4$.
Dual conformal invariance then uniquely fixes
the remaining 4-brackets.  
 This rule gives  all products $R_a R_b$ in terms of $\sixBra{1}{2}{3}{4}{5}{6}^4$.
Equation (\ref{six_point_lightlike_correlator_two_loops}) requires (\ref{NMHV_tree_ansatz_squared}) to be divided by the N${}^2$MHV tree-level amplitude, $\cAA{6}{2}^{(0)}$.  This is the anti-MHV amplitude~\eqref{antiMHV} which at six points is\footnote{Note that we use the tree-level anti-MHV superamplitude~\eqref{antiMHV} as input in our procedure.} 
\eq{\mathcal{A}_{6;\hspace{0.5pt}\text{N}^2\text{MHV}}^{\text{tree}}=\cAA{6}{2}^{(0)}= \frac{\sixBra{1}{2}{3}{4}{5}{6}^4}{\fourBra{1}{2}{3}{4} \fourBra{2}{3}{4}{5} \fourBra{3}{4}{5}{6} \fourBra{4}{5}{6}{1} \fourBra{5}{6}{1}{2} \fourBra{6}{1}{2}{3}}. \label{N2MHV_six_definition}}
As an example, consider the product  $R_1R_2$ \vspace{-0.5pt}
\eq{\begin{aligned}
R_1R_2&=  \frac{\fiveBra{2}{3}{4}{5}{6}^4 \fiveBra{3}{4}{5}{6}{1}^4}{ \fourBra{2}{3}{4}{5} \fourBra{3}{4}{5}{6} \fourBra{4}{5}{6}{2} \fourBra{5}{6}{2}{3} \fourBra{6}{2}{3}{4} \fourBra{3}{4}{5}{6} \fourBra{4}{5}{6}{1} \fourBra{5}{6}{1}{3} \fourBra{6}{1}{3}{4} \fourBra{1}{3}{4}{5}} \\[0.2ex]
&= \frac{\sixBra{1}{2}{3}{4}{5}{6}^4 \fourBra{3}{4}{5}{6}^4}{\fourBra{2}{3}{4}{5} \fourBra{3}{4}{5}{6} \fourBra{4}{5}{6}{2} \fourBra{5}{6}{2}{3} \fourBra{6}{2}{3}{4} \fourBra{3}{4}{5}{6} \fourBra{4}{5}{6}{1} \fourBra{5}{6}{1}{3} \fourBra{6}{1}{3}{4} \fourBra{1}{3}{4}{5}}, \end{aligned} \notag} 
where in the second line, the amplituhedron rule (\ref{amplituhedron_rule_one}) was used.

Proceeding in a similar way for  all other products in~\eqref{NMHV_tree_ansatz_squared}, we obtain simple rules for all products of $R$ invariants divided by the anti-MHV tree-level	 amplitude~\eqref{N2MHV_six_definition} in terms of ordinary bosonic twistor brackets:  
\eq{\begin{aligned}
		\frac{R_1R_2}{\mathcal{A}_{6;\hspace{0.5pt}\text{N}^2\text{MHV}}^{\text{tree}}} &=\frac{\fourBra{1}{2}{3}{4} \fourBra{1}{2}{3}{6} \fourBra{1}{2}{5}{6}\fourBra{3}{4}{5}{6}^3}{\fourBra{1}{3}{4}{5}\fourBra{1}{3}{4}{6}\fourBra{1}{3}{5}{6}\fourBra{2}{3}{4}{6}\fourBra{2}{3}{5}{6}\fourBra{2}{4}{5}{6}}, \\[0.1ex]
		\frac{R_1R_3}{\mathcal{A}_{6;\hspace{0.5pt}\text{N}^2\text{MHV}}^{\text{tree}}} &=\frac{\fourBra{1}{2}{3}{4} \fourBra{1}{2}{3}{6} \fourBra{2}{4}{5}{6}^2}{\fourBra{1}{2}{4}{5}\fourBra{1}{2}{4}{6}\fourBra{2}{3}{4}{6}\fourBra{2}{3}{5}{6}},\\[0.1ex]
		\frac{R_1R_4}{\mathcal{A}_{6;\hspace{0.5pt}\text{N}^2\text{MHV}}^{\text{tree}}} &=\frac{\fourBra{1}{2}{3}{4} \fourBra{1}{4}{5}{6} \fourBra{2}{3}{5}{6}^2}{\fourBra{1}{2}{3}{5}\fourBra{1}{3}{5}{6}\fourBra{2}{3}{4}{6}\fourBra{2}{4}{5}{6}}, \label{R_invariant_twistor_dictionary}
		\raisetag{1\baselineskip} 
\end{aligned}}
together with cyclic permutations of these.

Plugging these products into the ansatz for the square of the NMHV amplitude~\eqref{NMHV_tree_ansatz_squared} and  then into the duality equation~\eqref{six_point_lightlike_correlator_two_loops}, we equate the resulting expression\footnote{To avoid complicated twistor bracket identities, one can either do this  numerically or by rewriting twistor brackets in terms of $\tilde{z}_a-\tilde{z}_b$ via the relation $\fourBra{a}{b}{c}{d} = \varepsilon_{abcdef} (\tilde{z}_e-\tilde{z}_f) $, where $\tilde{z}_a \in \mathbb{C}$. This co-ordinate change was first used in \cite{1006.5703}.}
 to the known correlator~\eqref{six_point_lightlike_correlator_two_loops_x} (with the replacement $\x{a}{b}\rightarrow \fourBra{a{-}1}{\,a}{\,b{-}1}{\,b}$, see \eqref{xij2totwbr}). 
%

The resulting system of equations has  the following solution: \vspace{-4pt}
\eq{\alpha_1=\alpha_3=\alpha_5=\pm 1, \qquad  \alpha_2=\alpha_4=0, \label{tree_constraints}} 
so that 
\eq{\cAA{6}{1}^{(0)} = \pm( R_1+R_3+R_5).\label{pred}}
Thus we have derived the NMHV six-point tree-level amplitude from the 4-point correlator up to  an overall sign. Both signs yield the desired result for the correlator
\eq{
\frac{\left(\cAA{6}{1}^{(0)}\right)^2}{\mathcal{A}_{6;\hspace{0.5pt}\text{N}^2\text{MHV}}^{\text{tree}}}
=2
\left(\frac{\x{1}{3}\x{4}{6}}{\x{1}{4}\x{3}{6}}+\frac{\x{1}{5}\x{2}{4}}{\x{1}{4}\x{2}{5}}+\frac{\x{2}{6}\x{3}{5}}{\x{2}{5}\x{3}{6}}\right). \notag}
The known result is indeed given by~\eqref{pred} with the positive sign choice~\cite{0807.1095}.
This sign can clearly never be predicted purely by the correlator since the procedure predicts the square of the amplitude. If on the other hand we choose the wrong sign at tree level, this error will persist at higher loops and we will obtain the entire NMHV amplitude to all loops but with the wrong sign.

\vspace{-2pt}\subsection{One Loop}\label{subsec:prescriptive_six_point}\vspace{0pt}

At  one loop, the duality  (\ref{six_point_lightlike_correlator}) reads:
\eq{\lim_{\substack{\text{6-gon}\\\text{light-like}}}\!\! \xi^{(6)}\mathcal{F}^{(3)}=
	{\cAA{6}{0}^{(1)}}+\frac{\cAA{6}{2}^{(1)}}{\cAA{6}{2}^{(0)}} +\frac{\cAA{6}{1}^{(1)}\,\cAA{6}{1}^{(0)}}{\cAA{6}{2}^{(0)}}.\label{six_point_lightlike_correlator_three_loops}}
The first two terms form the MHV amplitude plus its parity conjugate whilst the last term is  a product of NMHV tree- and one-loop amplitudes.\footnote{This calculation differs from that of equation (4.19) in \cite{1103.4353} where the 5-point 2-loop correlator  was studied and the NMHV amplitude was given linearly.} 

As mentioned in the introduction, in order to go beyond tree level we require a basis of integrands.
At one loop we have the following basis of 23 independent planar boxes and parity-odd pentagons:
\vspace{-0.5pt}\begin{align}
	\cI_1^{(1)}&= \frac{\x{1}{3} \x{2}{4}}{\x{1}{\ell}\x{2}{\ell}\x{3}{\ell}\x{4}{\ell}} & &\text{one mass (6)}\notag\\
	\cI_7^{(1)}&= \frac{\x{1}{3} \x{2}{5}}{\x{1}{\ell}\x{2}{\ell}\x{3}{\ell}\x{5}{\ell}} & &\text{two-mass hard (6)}\notag\\
	\cI_{13}^{(1)}&=\frac{\x{1}{4} \x{2}{5}}{\x{1}{\ell}\x{2}{\ell}\x{4}{\ell}\x{5}{\ell}} & &\text{two-mass easy (3)}\notag\\
	\cI_{16}^{(1)}&=\frac{\x{1}{5} \x{2}{4}}{\x{1}{\ell}\x{2}{\ell}\x{4}{\ell}\x{5}{\ell}} & &\text{two-mass easy (3)}\notag\\
	\cI_{19}^{(1)}&=\frac{i \hspace{0.5pt}\varepsilon_{12345\ell} }{\x{1}{\ell}\x{2}{\ell}\x{3}{\ell}\x{4}{\ell}\x{5}{\ell}} & &\text{parity-odd pentagon (5)}
	\vspace{-0.5pt} \end{align}
where the list is understood to include all those related by cycling the six external variables (the numbers of independent integrands in each class is given in parentheses after each).
Note that there are only 5 independent parity-odd pentagons rather than 6 that one would expect from cyclicity. This is because there is an identity of the form
\eq{\cI_{19}^{(1)}-\cI_{20}^{(1)}+\cI_{21}^{(1)}-\cI_{22}^{(1)}+\cI_{23}^{(1)}-\cI_{24}^{(1)}=0,}
which we use to solve for $\cI_{24}^{(1)}$ in terms of the others. This identity is easily understood in the 6d embedding formalism where it can be written as
\eq{\frac{\varepsilon_{[L12345} X_{6]} \!\cdot \! X_L}{(X_1 \!\cdot \! X_L)(X_2 \!\cdot \! X_L)(X_3 \!\cdot \! X_L)(X_4 \!\cdot \! X_L)(X_5 \!\cdot \! X_L)(X_6 \!\cdot \! X_L)} =0.\label{pentid}}
Here the square bracket indicates anti-symmetrisation over 7 variables which yields zero in 6 dimensions. 
Our one-loop ans\"atze (see~\eqref{ansatz}) for the amplitudes thus reads
\eq{
	\mathcal{A}_{6;\hspace{0.2pt}0}^{(1)}=\sum_{j=1}^{23} a_j \hspace{0.5pt}\cI^{(1)}_j, \qquad \mathcal{A}_{6;\hspace{0.2pt}1}^{(1)}=\sum_{i=1}^5\sum_{j=1}^{23}  b_{ij}  \hspace{0.5pt} R_i\hspace{0.5pt} \cI^{(1)}_j, \qquad \mathcal{A}_{6;\hspace{0.2pt}2}^{(1)}=\mathcal{A}_{6;\hspace{0.2pt}2}^{(0)}\sum_{j=1}^{23} c_j \hspace{0.5pt}\cI^{(1)}_j.  \label{oneloop6ptans}}
The problem now involves solving a system of equations for the $23\!\times \!(1\hspace{-1pt}+\hspace{-1pt}5\hspace{-1pt}+\hspace{-1pt}1)\!=\! 161$ coefficients obtained by plugging these ans\"atze together with the previously found tree-level result~\eqref{pred} into~\eqref{six_point_lightlike_correlator_three_loops}. We will require the  products of $R$ invariants~\eqref{R_invariant_twistor_dictionary}. 
Moreover, we can use parity and cyclicity to immediately reduce the number of free coefficients.

\mbox{Equation (\ref{six_point_lightlike_correlator_three_loops})} can be evaluated at generic kinematic configurations. The {\sc Mathematica} package in \cite{1505.05886} generates convenient configurations of small magnitude in  random rational numbers. This process is repeated many times yielding a quadratic system over the rational numbers.

{	 Solving the system of equations with 161 coefficients arising from  (\ref{six_point_lightlike_correlator_three_loops}) we obtain  a solution with  $23$ free coefficients.  
	 Remarkably, the NMHV sector is entirely (and correctly) solved 
consistent with the comment in~footnote~\ref{footnote3}.
	  The anti-MHV sector is then fixed in terms of the MHV sector which is itself completely unfixed (hence 23 free coefficients---one for each integrand) and consistent with the ambiguity~\eqref{amb}. 
Imposing parity invariance, which takes $\mathcal{A}_{6;\hspace{0.2pt}0} \leftrightarrow \mathcal{A}_{6;\hspace{0.2pt}2}$ then reduces the number of free coefficients down to 5---the number of parity-odd integrands. Further imposing cyclicity reduces this down to just 1 free coefficient.
}

The resulting solution can be written as
\eq{\begin{aligned}
2\mathcal{A}_{6;\hspace{0.2pt}0}^{(1)}&=\sum_{j=1}^{6} \cI^{(1)}_j + \sum_{j=13}^{15} \big(\cI^{(1)}_j-\cI^{(1)}_{j+3}\big)- \alpha \big( \cI_{19}^{(1)} +\cI_{21}^{(1)}+\cI_{23}^{(1)}\big) \\
2\mathcal{A}_{6;\hspace{0.2pt}2}^{(1)}&=\left(\sum_{j=1}^{6} \cI^{(1)}_j + \sum_{j=13}^{15} \big(\cI^{(1)}_j-\cI^{(1)}_{j+3}\big) + \alpha \big( \cI_{19}^{(1)} +\cI_{21}^{(1)}+\cI_{23}^{(1)}\big)\right) \mathcal{A}_{6;\hspace{0.2pt}2}^{(0)} \\ 2\mathcal{A}_{6;\hspace{0.2pt}1}^{(1)}&=R_1\Big( \cI_3^{(1)}+\cI_6^{(1)} +\cI_8^{(1)}+\cI_{11}^{(1)}  +\frac13\big( \cI_{20}^{(1)} +\cI_{21}^{(1)}- \cI_{23}^{(1)}-\cI_{24}^{(1)}\big)  \Big) + \text{cyclic}.\label{oneloopsol}
\end{aligned}}
In the (anti-)MHV sector, we recognise the well known 1-loop result of a sum over one-mass and two-mass easy boxes together with an as yet undetermined parity-odd sector. The NMHV amplitude on the other hand is completely determined in terms of one-mass, two-mass hard boxes and parity-odd pentagons.

This prediction~\eqref{oneloopsol} agrees precisely with the known answer for $\alpha=1$. We will return to this as yet undetermined parameter $\alpha$ in the next subsection.

\vspace{-2pt}\subsection{Two Loops}\vspace{0pt}

We now proceed to two loops, using as input the  one-loop solution obtained above~\eqref{oneloopsol}.
We first need a basis of two-loop integrals.  A natural basis purely in position space is provided by dual conformal parity-even planar double boxes, pentaboxes, and pentapentagons, with all possible numerators, together with parity-odd pentaboxes and pentapentagons involving the 6d $\varepsilon$-tensor.

However, a convenient alternative dual conformal basis has been provided (together with an associated {\sc Mathematica} package) in~\cite{1505.05886,1704.05460} called the prescriptive basis. Although originally given in twistor space, all elements of this two-loop prescriptive basis can be rewritten in dual momentum space in terms of the planar basis described in the previous paragraph.
We attach this  $x$-space translation as a file to the work's {\tt arXiv} submission.

The prescriptive basis at two loops consists of $87$ elements.
These integrands we simply label as $\cI_i^{(2)}$ with $i=1,\ldots,87$.

We now insert the ans\"atze for the two-loop six-point amplitudes
\eq{\mathcal{A}_{6;\hspace{0.2pt}0}^{(2)}=\sum_{j=1}^{87} a_j \hspace{0.5pt} \cI^{(2)}_j, \qquad \mathcal{A}_{6;\hspace{0.2pt}1}^{(2)}=\sum_{i=1}^5\sum_{j=1}^{87}  b_{ij} \hspace{0.5pt} R_i \hspace{0.5pt} \cI^{(2)}_j, \qquad \mathcal{A}_{6;\hspace{0.2pt}2}^{(2)}=\mathcal{A}_{6;\hspace{0.25pt}2}^{(0)}\sum_{j=1}^{87} c_j \hspace{0.5pt} \cI^{(2)}_j ,}
comprising  of $87\! + \!5\! \times \!87\! +\! 87 \!=\! 609$ free coefficients, into the duality formula~\eqref{six_point_lightlike_correlator} which at this loop level reads
\eq{\lim_{\substack{\text{6-gon}\\\text{light-like}}}\!\! \xi^{(6)}\mathcal{F}^{(4)}=\cAA{6}{0}^{(2)}+\frac{\cAA{6}{2}^{(2)}}{\cAA{6}{2}^{(0)}}+\frac{\cAA{6}{0}^{(1)}\,\cAA{6}{2}^{(1)}}{\cAA{6}{2}^{(0)}} +\frac{\cAA{6}{1}^{(2)}\,\cAA{6}{1}^{(0)}}{\cAA{6}{2}^{(0)}}+\frac{1}{2}\frac{\big(\cAA{6}{1}^{(1)}\big)^2}{\cAA{6}{2}^{(0)}} .\label{six_point_lightlike_correlator_four_loops}}
Like the one-loop case, the whole NMHV sector at two loops is completely fixed by this equation. 
There are   
$87$ free undetermined coefficients in total, the anti-MHV sector being completely fixed in terms of the MHV sector, but the MHV sector itself being completely unfixed. This is precisely as expected in~\eqref{amb} and the accompanying footnote.
 Imposing parity then reduces the number of free coefficients to 36---the number of parity-odd two-loop planar dual conformal integrands. Further imposing cyclicity reduces this down  to 6---the number of cyclic classes of parity-odd integrands.
We expect these to be determined at the next loop order and cannot see any obstructions going to higher order.

The equations also (almost) determine the value of $\alpha$ in~\eqref{oneloopsol}:  the ambiguity at one loop.  
The equations are clearly quadratic in one-loop parameters and in fact, this  gives rise to \textit{two} possible solutions. This is evident as the correlator determines only the parity-symmetric product
\eq{\frac{\mathcal{A}_{6;\hspace{0.2pt}0}^{(1)} \, \mathcal{A}_{6;\hspace{0.2pt}2}^{(1)}}{\mathcal{A}_{6;\hspace{0.2pt}2}^{(0)}}= \big(\mathcal{M}_{6}^{(1)}\big\vert_{\text{even}}\big)^2 \hspace{-1.5pt} -\hspace{-1.5pt}  \big(\mathcal{M}_{6}^{(1)}\big\vert_{\text{odd}}\big)^2 , \label{MHV1_MHVbar1}}
for $\mathcal{M}_{6}^{(1)}\big\vert_{\text{even}}\hspace{-1.5pt}= \!\big(\mathcal{M}_6^{(\ell)}\hspace{-1.3pt}+\hspace{-1.3pt}\overline{\mathcal{M}}_6^{(\ell)}\big)/2$ and $\mathcal{M}_{6}^{(1)}\big\vert_{\text{odd}}\hspace{-1.5pt}=\!\big(\mathcal{M}_6^{(\ell)}\hspace{-1.3pt}-\hspace{-1.3pt}\overline{\mathcal{M}}_6^{(\ell)}\big)/2$ where $\mathcal{M}_6$, $\overline{\mathcal{M}}_6$ are the MHV, anti-MHV amplitudes normalised by their respective tree-level amplitudes (so 
 $\mathcal{M}_6=\mathcal{A}_{6;\hspace{0.2pt}0}^{(1)}$, $\overline{\mathcal{M}}_6=\mathcal{A}_{6;\hspace{0.2pt}2}^{(1)}/\mathcal{A}_{6;\hspace{0.2pt}2}^{(0)}$). The even piece was determined at one loop whilst $(\mathcal{M}_{6}^{(1)}\big\vert_{\text{odd}})^2$ is determined by this equation. This yields $\alpha^2\hspace{-2pt}=\hspace{-2pt}1$ so $\alpha\hspace{-2pt}=\hspace{-2pt}\pm 1$. We thus see that this procedure alone cannot resolve the sign of the parity-odd part at one loop.
This additional sign ambiguity is only present for MHV amplitudes and is a purely one-loop effect.
Note that the ambiguity simply interchanges the MHV and anti-MHV solutions.

As a final note, the resulting integrand is consistent with that obtained in~\cite{1505.05886} and can be retrieved explicitly via the associated {\sc Mathematica} package.

\vspace{-6pt}\section{Seven-Point Integrands}\label{sec:extracting_seven_point_integrands}\vspace{-4pt}
In this section, we study the seven-point light-like limit of the correlator, continuing our extraction of amplitudes from the correlator. The construction now involves the null separation of seven adjacent points. The statement of the duality from (\ref{n_point_duality}) is
\eq{\lim_{\substack{\text{7-gon}\\\text{light-like}}} \Big(  \xi^{(7)}\mathcal{F}^{(\ell+3)}\Big)
	=\sum_{m=0}^{\ell} \frac{\mathcal{A}_{7;\hspace{0.2pt}0}^{(m)}\,\mathcal{A}_{7;\hspace{0.2pt}3}^{(\ell-m)}    +\mathcal{A}_{7;\hspace{0.2pt}1}^{(m)}  \mathcal{A}_{7;\hspace{0.2pt}2}^{(\ell-m)}   }{\mathcal{A}_{7;\hspace{0.2pt}3}^{(0)}},\label{seven_point_lightlike_correlator}}
where all amplitudes are normalised by the tree-level MHV amplitude.

Just like six points, we will  proceed order-by-order  in the coupling, making amplitude integrand predictions from the correlator. To do so, we first require an understanding of the building blocks involved. In particular, we need to understand the N${}^2$MHV  super-invariants at seven points and how to multiply these with NMHV $R$ invariants.
\vspace{-2pt}\subsection{The Yangian Invariants}\label{subsec:seven_point_integrands_tree}\vspace{0pt}
The tree-level anti-MHV (=N${}^3$MHV) amplitude is (\ref{antiMHV})
\eq{\mathcal{A}_{7;\hspace{0.5pt}\text{N}^3\text{MHV}}^{\text{tree}}=\cAA{7}{3}^{(0)}= \frac{\langle 1\hspace{0.5pt}2\hspace{0.5pt}3\hspace{0.5pt}4\hspace{0.5pt}5\hspace{0.5pt}6\hspace{0.5pt}7 \rangle^4}{\fourBra{1}{2}{3}{4}\fourBra{2}{3}{4}{5}\fourBra{3}{4}{5}{6}\fourBra{4}{5}{6}{7}\fourBra{5}{6}{7}{1}\fourBra{6}{7}{1}{2}\fourBra{7}{1}{2}{3}}. \label{N3MHV_seven_definition}}
At the NMHV level, we assume an expansion of the amplitude in terms of $R$ invariants.
Let us define a short-hand notation for the seven-point $k\hspace{-1.5pt}=\hspace{-1.5pt}1$ $R$ invariants~\eqref{R_invariant_definition}:
\eq{R_{7;\hspace{0.5pt}(a),(b)}^{(k=1)}\equiv \RInv{c}{d}{e}{f}{g}, \label{R_invariant_seven_definition}}
which is just the ordered $R$ invariant involving external points $c,d,e,f,g$ with $a, b$ missing. In fact, this notation is very natural from the point of view of the Grassmannian: the $R$ invariant  $R_{7;\hspace{0.5pt}(a),(b)}^{(k=1)}$ corresponds to the residue of the relevant Grassmannian integral ($\Gr(1,7)$) evaluated at the poles $(a)=0, (b)=0$. 

There are clearly 21 of these $R$ invariants, however they are not all independent.
The identities the $R$ invariants satisfy arise from the six-point identity~\eqref{Rid}, namely for any six points
\eq{\RInv{a}{b}{c}{d}{e}-\RInv{a}{b}{c}{d}{f}+\RInv{a}{b}{c}{e}{f}-\RInv{a}{b}{d}{e}{f}+\RInv{a}{c}{d}{e}{f}-\RInv{b}{c}{d}{e}{f}=0.
\label{Rid2}}
At seven points, there are 7 such identities, but only 6 of them are in fact independent. We are therefore left with $21-6=15$ independent $R$ invariants.

The $\text{N}^2$MHV sector however requires more thought. We follow the procedure outlined in subsection~\ref{subsec:Yangian_invariants_from_Grass} to obtain N${}^2$MHV Yangian invariants in amplituhedron space from the Grassmannian.
We illustrate this for the simplest example and provide the ingredients for every other seven-point residue in Appendix \ref{appendix_covariantisation}.

Recall from \mbox{equation (\ref{grassmannian_integral})} that any 7-point N${}^k$MHV Yangian invariant can be represented as the Grassmannian integral
\eq{
	\frac{1}{\text{vol}\hspace{0.8pt}[\GL(2)]} \int \frac{d^{\hspace{1pt}2\times 7} C_{\alpha a}}{(1 2)(23)(34)(45)(56)(67)(71)}\prod_{\alpha=1}^2\delta^{4|4}(C_{\alpha a}\cZ_{a}^{\mathcal{A}}). \label{grassmannian_integral_seven}}
The integration is 10 dimensional  (after dividing by the four-dimensional vol[$\GL(2)$]) and there are 8 bosonic delta functions, leaving 2 non-trivial integrations. 
These we can choose to circle two poles and use the residue theorem. 

There are three classes of residues from the following vanishing minors
\eq{
(67)\hspace{-1.5pt}=\hspace{-1.5pt}(71)\hspace{-1.5pt}=\hspace{-1.5pt}0,\qquad
(12)\hspace{-1.5pt}=\hspace{-1.5pt}(34)\hspace{-1.5pt}=\hspace{-1.5pt}0, \qquad
(12)\hspace{-1.5pt}=\hspace{-1.5pt}(45)\hspace{-1.5pt}=\hspace{-1.5pt}0, \label{grassmannian_seven_minors} } 
where all other invariants are related by cyclicity. The simplest case is the residue at the pole $(67)\hspace{-1.5pt}=\hspace{-1.5pt}(71)\hspace{-1.5pt}=\hspace{-1.5pt}0$. We can pick
\textit{canonical} positive co-ordinates on the Grassmannian restricted to this subspace, as found in \cite{1212.5605}
\eq{
C_{\alpha a}=\begin{bmatrix} 1 & \alpha_2\hspace{-1.5pt}+\hspace{-1.5pt}\alpha_4\hspace{-1.5pt}+\hspace{-1.5pt}\alpha_6\hspace{-1.5pt}+\hspace{-1.5pt}\alpha_8 & (\alpha_2\hspace{-1.5pt}+\hspace{-1.5pt}\alpha_4\hspace{-1.5pt}+\hspace{-1.5pt}\alpha_6)\alpha_7 & (\alpha_2\hspace{-1.5pt}+\hspace{-1.5pt}\alpha_4)\alpha_5 & \alpha_2\alpha_3 & 0 & 0 \\
0 & 1 & \alpha_7 & \alpha_5 & \alpha_3 & \alpha_1 & 0
\end{bmatrix},}
for which the (residue of the) measure  of the Grassmannian integral becomes
\eq{\Omega=	\int \frac{\text{d}\alpha_1 \dots \text{d}\alpha_8}{\alpha_1 \dots \alpha_8} .}
From~\eqref{Y_weightless_form}, we can then jump straight to the Yangian invariant in amplituhedron space by solving
\eq{\Omega = \frac{\text{d}\alpha_1 \dots \text{d}\alpha_8}{\alpha_1 \dots \alpha_8} = \twoBra{Y}{d^{\hspace{0.5pt}4} Y_1} \twoBra{Y}{d^{\hspace{0.5pt}4} Y_2} \times \mathcal{Y}_{7;\hspace{0.2pt}2}(\hat Z_1,\dots,\hat Z_7, Y), \label{eq}}
where $Y=C_{\alpha a} \hat Z_{a}^{{\tilde A}}$.  Using $\GL(6)$ invariance, we can choose amplituhedron co-ordinates as
\eq{\hat Z_{a}^{\tilde{A}}  =\begin{bmatrix}1 & 0 & 0 & 0 & 0 & 0 & A \\[-0.5ex]
0 & 1 & 0 & 0 & 0 & 0 & B \\[-0.5ex]
0 & 0 & 1 & 0 & 0 & 0 & C \\[-0.5ex]
0 & 0 & 0 & 1 & 0 & 0 & D \\[-0.5ex]
0 & 0 & 0 & 0 & 1 & 0 & E \\[-0.5ex]
0 & 0 & 0 & 0 & 0 & 1 & F \\
\end{bmatrix},\label{coords}}
giving 
\eq{Y_{\alpha}^{\tilde{A}}=\begin{bmatrix} 1 & \alpha_2\hspace{-1.5pt}+\hspace{-1.5pt}\alpha_4\hspace{-1.5pt}+\hspace{-1.5pt}\alpha_6\hspace{-1.5pt}+\hspace{-1.5pt}\alpha_8 & (\alpha_2\hspace{-1.5pt}+\hspace{-1.5pt}\alpha_4\hspace{-1.5pt}+\hspace{-1.5pt}\alpha_6)\alpha_7 & (\alpha_2\hspace{-1.5pt}+\hspace{-1.5pt}\alpha_4)\alpha_5 & \alpha_2\alpha_3 & 0  \\
	0 & 1 & \alpha_7 & \alpha_5 & \alpha_3 & \alpha_1 
\end{bmatrix},}
which in turn yields
\eq{\begin{aligned}\twoBra{Y}{d^4 Y_1}\twoBra{Y}{d^4 Y_2} 
		= \alpha_1 \alpha_3 \alpha_5 \alpha_7 \hspace{1.5pt} \text{d}\alpha_1 \dots \text{d}\alpha_8. \vspace{-10pt} \label{Y_form_6771}\end{aligned}} 
The differential form is clearly weight $6$ in $Y$ giving us the freedom to divide by \textit{any} six brackets to obtain a $Y$-weightless volume form, let us choose:
\eq{\frac{\twoBra{Y}{d^4 Y_1}\twoBra{Y}{d^4 Y_2}}{\fiveBra{Y}{1}{2}{3}{4} \fiveBra{Y}{1}{2}{3}{6}  \fiveBra{Y}{1}{4}{5}{6} \fiveBra{Y}{2}{3}{4}{5} \fiveBra{Y}{2}{3}{4}{6} \fiveBra{Y}{3}{4}{5}{6}} =  \frac{\text{d}\alpha_1 \dots \text{d}\alpha_8}{\alpha_1\alpha_2\alpha_3^2\alpha_4 \alpha_8}.  \label{Y_form_weightless_6771}} 
Therefore, the $(67)=(71)=0$ residue is given as
\eq{\Omega_{(67),(71)}\equiv \frac{\text{d}\alpha_1 \dots \text{d}\alpha_8}{\alpha_1\dots\alpha_8}=\frac{\alpha_3  \twoBra{Y}{d^4 Y_1}\twoBra{Y}{d^4 Y_2}}{\alpha_5 \alpha_6 \alpha_7 \hspace{1.5pt}\fiveBra{Y}{1}{2}{3}{4} \fiveBra{Y}{1}{2}{3}{6}  \fiveBra{Y}{1}{4}{5}{6} \fiveBra{Y}{2}{3}{4}{5} \fiveBra{Y}{2}{3}{4}{6} \fiveBra{Y}{3}{4}{5}{6}} .  \label{amplituhedron_6771_first}} 
We now wish to covariantise this expression.
To achieve this, we simply need covariant expressions for the Grassmannian co-ordinates---which are the following:
\vspace{3pt}
\eq{\begin{aligned}
&\alpha_1 = \frac{\fiveBra{Y}{2}{3}{4}{5}}{\fiveBra{Y}{3}{4}{5}{6}},\hspace{5pt}\alpha_2= -\frac{\fiveBra{Y}{1}{2}{3}{4}\fiveBra{Y}{3}{4}{5}{6}}{\fiveBra{Y}{2}{3}{4}{5}\fiveBra{Y}{2}{3}{4}{6}},\hspace{5pt}\alpha_3 = -\frac{\fiveBra{Y}{2}{3}{4}{6}}{\fiveBra{Y}{3}{4}{5}{6}},\hspace{5pt}\alpha_4= -\frac{\fiveBra{Y}{1}{2}{3}{6}\fiveBra{Y}{3}{4}{5}{6}}{\fiveBra{Y}{2}{3}{4}{6}\fiveBra{Y}{2}{3}{5}{6}},\\
&\alpha_5 = \frac{\fiveBra{Y}{2}{3}{5}{6}}{\fiveBra{Y}{3}{4}{5}{6}},\hspace{5pt}\alpha_6= -\frac{\fiveBra{Y}{1}{2}{5}{6}\fiveBra{Y}{3}{4}{5}{6}}{\fiveBra{Y}{2}{3}{5}{6}\fiveBra{Y}{2}{4}{5}{6}},\hspace{5pt}\alpha_7= -\frac{\fiveBra{Y}{2}{4}{5}{6}}{\fiveBra{Y}{3}{4}{5}{6}},\hspace{5pt}\alpha_8= -\frac{\fiveBra{Y}{1}{4}{5}{6}}{\fiveBra{Y}{2}{4}{5}{6}}.
\end{aligned} \label{grassmannian_coord_relations_6771}}
We require the above cross ratios to be $Y$-weightless, so that their combinations in (\ref{amplituhedron_6771_first}) are $Y$-weightless. Plugging these in yields
	\eq{\Omega_{(67),(71)}\rightarrow\frac{\twoBra{Y}{d^4 Y_1}\twoBra{Y}{d^4 Y_2}}{\fiveBra{Y}{1}{2}{3}{4} \fiveBra{Y}{2}{3}{4}{5}  \fiveBra{Y}{3}{4}{5}{6} \fiveBra{Y}{4}{5}{6}{1} \fiveBra{Y}{5}{6}{1}{2} \fiveBra{Y}{6}{1}{2}{3}}.  \label{amplituhedron_6771_second}} 
Whilst the expression is weightless in $Y$, the external particles are still weighted.
Although this is correct for the choice of co-ordinates~\eqref{coords}, we use the following ($Y$-weightless) relations
\eq{\begin{gathered}
A=-\sixBra{2}{3}{4}{5}{6}{7},\hspace{5pt}B=\sixBra{1}{3}{4}{5}{6}{7},\hspace{5pt}C=-\sixBra{1}{2}{4}{5}{6}{7},\\
D=\sixBra{1}{2}{3}{5}{6}{7},\hspace{5pt}E=-\sixBra{1}{2}{3}{4}{6}{7},\hspace{5pt}F=\sixBra{1}{2}{3}{4}{5}{7},\hspace{5pt} 1= \sixBra{1}{2}{3}{4}{5}{6},
\end{gathered} \label{grassmannian_capital_relations_6771}}
to obtain a co-ordinate independent result (in general, the result would depend non-trivially on the unfixed co-ordinates $A,B,\ldots$).
The natural modification here  
is simply to multiply by $\sixBra{1}{2}{3}{4}{5}{6}^4 = 1$ 
\eq{
	\Omega_{(67),(71)}\rightarrow
	\frac{
		\twoBra{Y}{d^4 Y_1}\twoBra{Y}{d^4 Y_2}
		\sixBra{1}{2}{3}{4}{5}{6}^4}
	{\fiveBra{Y}{1}{2}{3}{4} \fiveBra{Y}{2}{3}{4}{5}  \fiveBra{Y}{3}{4}{5}{6} \fiveBra{Y}{4}{5}{6}{1} \fiveBra{Y}{5}{6}{1}{2} \fiveBra{Y}{6}{1}{2}{3}},  \label{amplituhedron_6771_final}} 
which is the covariant expression for the desired residue.
This example is somewhat trivial and indeed could have been obtained by simply realising that the invariant is secretly the unique six-point N${}^2$MHV Yangian invariant. 

The other cases in~\eqref{grassmannian_seven_minors} are less trivial but can be computed using this same method.
 An outline for deriving these from the Grassmannian can be found in \mbox{Appendix \ref{appendix_covariantisation}} and we simply present them here:
\eq{\begin{aligned}
 		&R_{(67),(71)}^{(k=2)}\equiv\frac{\sixBra{1}{2}{3}{4}{5}{6}^4}{\fourBra{1}{2}{3}{4} \fourBra{2}{3}{4}{5}  \fourBra{3}{4}{5}{6} \fourBra{4}{5}{6}{1} \fourBra{5}{6}{1}{2} \fourBra{6}{1}{2}{3}},\\
 		&R_{(12),(34)}^{(k=2)}\equiv\frac{( \langle[ 1 | \hspace{0.5pt} 5 \hspace{0.5pt}  6  \hspace{0.5pt} 7 \rangle \langle | 2 ] \hspace{0.5pt} 3 \hspace{0.5pt} 4 \hspace{0.5pt}5 \hspace{0.5pt} 6 \hspace{0.5pt} 7 \rangle)^4}{\fourBra{1}{2}{6}{7}\fourBra{1}{5}{6}{7}\fourBra{2}{5}{6}{7}\fourBra{3}{4}{5}{6}\fourBra{3}{5}{6}{7}\fourBra{4}{5}{6}{7}
 			\langle  1 \hspace{0.5pt} 2 \hspace{0.5pt} 5 [ 7 | \rangle \langle3 \hspace{0.5pt} 4 \hspace{0.5pt} 5 | 6 ] \rangle \langle  1 \hspace{0.5pt} 2  \hspace{0.5pt} [ 6 | 7  \rangle \langle  3 \hspace{0.5pt} 4  | 5 ] \hspace{0.5pt} 7 \rangle},\\
 		&R_{(12),(45)}^{(k=2)}\equiv\frac{( \langle [ 2 | \hspace{0.5pt} 3 \hspace{0.5pt}  6  \hspace{0.5pt} 7 \rangle \langle | 1 ] \hspace{0.5pt} 3 \hspace{0.5pt} 4 \hspace{0.5pt}5 \hspace{0.5pt} 6 \hspace{0.5pt} 7 \rangle)^4}{\fourBra{1}{2}{3}{7}\fourBra{1}{2}{6}{7}\fourBra{1}{3}{6}{7}\fourBra{2}{3}{6}{7}\fourBra{3}{4}{5}{6}\fourBra{3}{4}{6}{7}\fourBra{3}{5}{6}{7}\fourBra{4}{5}{6}{7}
 			\langle  1 \hspace{0.5pt} 2 \hspace{0.5pt} 3 [ 7 | \rangle \langle 3 \hspace{0.5pt} 4 \hspace{0.5pt} 5 | 6 ] \rangle},
 	\end{aligned}\label{seven_point_N2MHV_yangian_invariants}}
 where for example, $\langle[ 1 | \hspace{0.5pt} 5 \hspace{0.5pt}  6  \hspace{0.5pt} 7 \rangle \langle | 2 ] \hspace{0.5pt} 3 \hspace{0.5pt} 4 \hspace{0.5pt}5 \hspace{0.5pt} 6 \hspace{0.5pt} 7 \rangle \hspace{-1pt} \equiv \hspace{-1pt}  \fourBra{1}{5}{6}{7}\sixBra{2}{3}{4}{5}{6}{7}\hspace{-1.5pt}-\hspace{-1.5pt}\fourBra{2}{5}{6}{7}\sixBra{1}{3}{4}{5}{6}{7} $ is an ordered anti-symmetrisation for two points enclosed in a square bracket.

These 21 N${}^2$MHV invariants are conjugates to the 21 NMHV $R$ invariants as follows
\eq{\begin{aligned}
		&\overline{\RInv{3}{4}{5}{6}{7}}=\overline{R}_{(1),(2)}^{(k=1)}=R_{(45),(56)}^{(k=2)},\\&\overline{\RInv{2}{4}{5}{6}{7}}=\overline{R}_{(1),(3)}^{(k=1)}=R_{(45),(67)}^{(k=2)},\\&\overline{\RInv{2}{3}{4}{6}{7}}=\overline{R}_{(1),(5)}^{(k=1)}={R}_{(45),(12)}^{(k=2)}.\end{aligned}\label{R1_R2_conjugation_relations}}
	These conjugation relations can be seen from  the Grassmannian. In complete generality, conjugation relates ordered minors in the $\Gr(k,n)$ Grassmannian to those of the conjugate Grassmannian $\Gr(n{-}k{-}4,n)$ as follows%
\footnote{There are two equivalent Grassmannian formulae for N${}^k$MHV amplitudes, the $\Gr(k,n)$ one we use here which manifests dual conformal symmetry,  and the $\Gr(k+2,n)$ one which manifests the original conformal symmetry. Conjugation is more transparent in the $\Gr(k+2,n)$ case where it takes $C \rightarrow C^\perp\in \Gr(n-k-2,n)$ where the minors are related via $(a,b,\dots, c)=\varepsilon_{a,b,..,c,d,e,..,f}(d,e,..,f)^\perp$.
The relation between ordered minors in $\Gr(k,n)$ and those  in $\Gr(k+2,n)$ is simply $\Gr(k,n)\ni(a\,a{+}1,\dots, a{+}k{-}1)=(a{-}1,a,\dots,a{+}k)\in\Gr(k+2,n)$~\cite{0909.0483}. From here we see the conjugation relation~\eqref{conj} for minors in the $\Gr(k,n)$ formalism we are considering.
}
\eq{(a,\,a{+}1,\dots, a{+}k{-}1)\qquad 
		\xrightarrow{\text{conjugation}} \qquad  (a{+}k{+}2,\,a{+}k{+}3,\dots, a{+}n{-}3). \label{conj}}
In the current context, conjugation takes the $k\!=\!1$ poles $(a)$ to the $k\!=\!2$ poles $(a{+}3,\,a{+}4)$. This then implies the corresponding relations between Yangian invariants~\eqref{R1_R2_conjugation_relations}.

With  these conjugation relations, we can immediately obtain  the N${}^2$MHV identities  which now follow directly from~\eqref{Rid2}. Just like the NMHV $R$ invariants, there are therefore 6 independent identities leaving 15 independent N${}^2$MHV invariants.

As well as the Yangian invariants themselves, we also need an understanding on how to take products of NMHV and N${}^2$MHV Yangians.
Again, this is essentially determined by considering the conformal weights, similarly to~\eqref{amplituhedron_rule_one}, namely if a six- and five-bracket have five points in common, this gives a vanishing result. The only other possibility at seven points  is that they have four points in common in which case we get
\eq{\sixBra{a}{b}{c}{d}{e}{f}^4 \fiveBra{a}{b}{c}{d}{g}^4= \sevenBra{a}{b}{c}{d}{e}{f}{g}^4 \fourBra{a}{b}{c}{d}^4. \label{amplituhedron_rule_twp}}
\newpage
\vspace{-2pt}\subsection{Tree Level}
\vspace{0pt}


We now proceed similarly to  six points: we first write down an ansatz for the seven-point NMHV (N${}^2$MHV) amplitudes as an arbitrary linear combination of the independent $k=1$ ($k=2$)  Yangian invariants (15 each)
\eq{\mathcal{A}_{7;\hspace{0.2pt}1}^{(0)}=\sum_{i=1}^{15}a_{i} \hspace{0.5pt} R^{(k=1)}_i, \qquad \qquad 
\mathcal{A}_{7;\hspace{0.5pt}2}^{(0)}=\sum_{i=1}^{15}b_{i} \hspace{0.2pt} R^{(k=2)}_i, \label{tree7ptans}}
where we list an arbitrary set of independent super-invariants (defined in the previous subsection) by $R_i^{(k)}$.

We then plug these ans\"atze into the duality formula~\eqref{seven_point_lightlike_correlator} which at tree level becomes
\eq{\lim_{\substack{\text{7-gon}\\\text{light-like}}}\!\! \xi^{(7)}\mathcal{F}^{(3)}=
	1+\frac{\cAA{7}{1}^{(0)}\,\cAA{7}{2}^{(0)}}{\cAA{7}{3}^{(0)}}.\label{seven_point_lightlike_correlator_three_loops}}
Using the formula for taking products~\eqref{amplituhedron_rule_twp} as well as the known N${}^3$MHV tree-level amplitude~\eqref{N3MHV_seven_definition} yields an algebraic equation in the 30 unknowns. For convenience, we provide explicit expressions for all the products of Yangian invariants in an attached {\sc Mathematica} notebook.

Again proceeding numerically, evaluating all twistor brackets at random rational points many times, 
one obtains a $1$-parameter solution---so far without imposing parity or cyclicity. This free parameter is an overall scaling of the NMHV amplitude with the inverse scaling of the N$^2$MHV sector, which the light-like correlator will not detect
\eq{\mathcal{A}_{7;\hspace{0.5pt}\text{N}\text{MHV}}\hspace{-1.5pt}\times \hspace{-1.5pt} \mathcal{A}_{7;\hspace{0.5pt}\text{N}^2\text{MHV}} = \alpha \mathcal{A}_{7;\hspace{0.5pt}\text{N}\text{MHV}} \hspace{-1.5pt} \times \hspace{-1.5pt} \frac{1}{\alpha}\mathcal{A}_{7;\hspace{0.5pt}\text{N}^2\text{MHV}}.}
However, imposing parity invariance  clearly fixes $\alpha=1/\alpha$, so that $\alpha=\pm 1$. We can therefore fix the tree-level amplitude up to a sign from the correlator; the result can be written (with the correct choice of sign):
\eq{\begin{aligned}\mathcal{A}_{7;\hspace{0.5pt}\text{N}\text{MHV}}^{\text{tree}}&=R_{(5), (6)}^{(k=1)} + R_{(6), (1)}^{(k=1)} + R_{(1), (2)}^{(k=1)} + 
	R_{(1), (4)}^{(k=1)} + R_{(3), (6)}^{(k=1)} + R_{(3), (4)}^{(k=1)},\notag\\[0.9pt]
	\mathcal{A}_{7;\hspace{0.5pt}\text{N}^2\text{MHV}}^{\text{tree}}&=R_{(1 2), (2 3)}^{(k=2)} + R_{(2 3), (4 5)}^{(k=2)} + R_{(4 5), (5 6)}^{(k=2)} + 
 R_{(4 5), (71)}^{(k=2)} + R_{(6 7), (2 3)}^{(k=2)} + R_{(6 7), (7 1)}^{(k=2)}. \label{N2MHV_seven_definition}
\end{aligned}}
Note that cyclicity was not input: the result is of course cyclically invariant although one has to use the identities to see this.

%
\vspace{-2pt}\subsection{One Loop}\label{subsec:seven_point_integrands_one_loop}\vspace{0pt}
We now wish to extract all seven-point one-loop amplitudes from the correlator. A complete basis of dual conformal one-loop integrands is given by the following parity-even integrands together with their 7 cyclic versions each
\vspace{-0.5pt} \begin{align}
	\cI_1^{(1)} &= {\x{1}{3} \x{2}{4} \over \x{1}{\ell} \x{2}{\ell} \x{3}{\ell} \x{4}{\ell}} &&\text{one mass}\notag\\
	 	\cI_8^{(1)} &= {\x{1}{3} \x{2}{5} \over \x{1}{\ell} \x{2}{\ell} \x{3}{\ell} \x{5}{\ell}}  &&\text{two-mass hard}\notag\\
\cI_{15}^{(1)} &= {\x{1}{3} \x{2}{6} \over \x{1}{\ell} \x{2}{\ell} \x{3}{\ell} \x{6}{\ell}}  &&\text{two-mass hard}\notag\\
\cI_{22}^{(1)} &= {\x{1}{4} \x{2}{5} \over \x{1}{\ell} \x{2}{\ell} \x{4}{\ell} \x{5}{\ell}}  &&\text{two-mass easy}\notag\\
\cI_{29}^{(1)} &= {\x{1}{5} \x{2}{4} \over \x{1}{\ell} \x{2}{\ell} \x{4}{\ell} \x{5}{\ell} }  &&\text{two-mass easy}\notag\\
\cI_{36}^{(1)} &= {\x{1}{4} \x{2}{6} \over \x{1}{\ell} \x{2}{\ell} \x{4}{\ell} \x{6}{\ell}     }  &&\text{three mass }\notag\\
\cI_{43}^{(1)} &= {\x{1}{6} \x{2}{4} \over \x{1}{\ell} \x{2}{\ell} \x{4}{\ell} \x{6}{\ell}   }  &&\text{three mass }\vspace{-0.5pt} \end{align}
giving 49 independent parity-even integrands in total.
There are also 21 parity-odd pentagons 
\eq{\cI_{abcde}^{(1)} = \frac{i \hspace{0.5pt} \varepsilon_{abcde\ell}}{\x{a}{\ell}\x{b}{\ell}\x{c}{\ell}\x{d}{\ell}\x{e}{\ell}}.
}
These parity-odd pentagons satisfy identities which follow directly  from~\eqref{pentid}. Amusingly, these are  exactly the  same six-term identity that the NMHV $R$ invariants $\RInv{a}{b}{c}{d}{e}$ satisfy, thus there are 15 independent parity-odd integrands (the same number as independent $R$ invariants).
In total, there are $49\! + \!15\!=\!64$ independent one-loop integrands at seven points.

So the ans\"atze for the one-loop amplitudes at seven points are
\vspace{-0.5pt}\begin{align}
\mathcal{A}_{7;\hspace{0.2pt}0}^{(1)}&=\sum_{j=1}^{64} a_j \hspace{0.5pt} \cI^{(1)}_j, \qquad& \mathcal{A}_{7;\hspace{0.2pt}1}^{(1)}&=\sum_{i=1}^{15}\sum_{j=1}^{64}  b_{ij} \hspace{0.2pt}  R^{(k=1)}_i \hspace{0.5pt}\cI^{(1)}_j, \notag\\
\mathcal{A}_{7;\hspace{0.2pt}2}^{(1)}&=\sum_{i=1}^{15}\sum_{j=1}^{64}  c_{ij}  \hspace{0.2pt} R^{(k=2)}_i \hspace{0.5pt} \cI^{(1)}_j,
\qquad& \mathcal{A}_{7;\hspace{0.2pt}3}^{(1)}&=\mathcal{A}_{7;\hspace{0.2pt}3}^{(0)}\sum_{j=1}^{64} d_j \hspace{0.5pt}\cI^{(1)}_j , \label{oneloop7ptans}
\vspace{-0.5pt}\end{align}
with $64\!\times\!(1{+}15{+}15{+}1)=2048$ coefficients.

The correlator/amplitude duality at this order gives
\eq{\lim_{\substack{\text{7-gon}\\\text{light-like}}}\!\! \xi^{(7)}\mathcal{F}^{(4)}=\cAA{7}{0}^{(1)}+\frac{\cAA{7}{3}^{(1)}}{\cAA{7}{3}^{(0)}} +\frac{\cAA{7}{1}^{(1)}\,\cAA{7}{2}^{(0)}}{\cAA{7}{3}^{(0)}}+\frac{\cAA{7}{2}^{(1)}\,\cAA{7}{1}^{(0)}}{\cAA{7}{3}^{(0)}}.\label{seven_point_lightlike_correlator_four_loops}}
Plugging in the above ans\"atze and using the product rule between $k\!=\!1$ and $k\!=\!2$ invariants \eqref{amplituhedron_rule_twp} gives a set of equations for the coefficients in terms of twistor brackets.

Solving the  resulting equation (numerically using random rationals for the twistors), we obtain a solution with 128 free coefficients. This is precisely as expected from the general discussion of~\eqref{amb}; there is an ambiguity of both the MHV and NMHV amplitude in the form of the tree-level amplitude times any combination of the 64 one-loop integrands.
The N${}^2$MHV and N${}^3$MHV amplitudes are then fixed in terms of these.

Parity reduces the solution down to  $15{+}15\!=\!30$ coefficients---the ambiguity projects to only parity-odd integrands. Applying cyclicity in addition reduces this down to $3{+}3\!=\!6$ free  coefficients, corresponding to the 3 cyclic  classes of parity-odd integrands for both MHV and NMHV sectors.

\vspace{-2pt}\subsection{Two Loops}
\vspace{0pt}

Finally, we proceed to two loops expecting to fix the remaining one-loop coefficients as well as determining the parity-even part of the two-loop answer.

As for six points, the two-loop basis consists of all dual conformal double boxes, pentaboxes and pentapentas, built either from $\x{a}{b}$ only (parity-even) or in addition, a single six-dimensional $\varepsilon$-tensor.
Just like six points, we again find it convenient to use the smaller prescriptive basis of~\cite{1505.05886} and the accompanying package.  
These were all originally given in terms of twistor brackets, but can all be converted to an $x$-space representation where they are all linear combinations of this dual conformal $x$-basis.
We provide the result of this translation explicitly in a file attached to the {\tt arXiv} submission. There are 378 integrands in the two-loop seven-point prescriptive basis which we label here as $\cI_i^{(2)}$. 
So we have the  following ans\"atze for the two-loop amplitudes
\vspace{-0.5pt}\begin{align}
\mathcal{A}_{7;\hspace{0.2pt}0}^{(2)}&=\sum_{j=1}^{378} a_j \hspace{0.5pt} \cI^{(2)}_j, \qquad& \mathcal{A}_{7;\hspace{0.2pt}1}^{(2)}&=\sum_{i=1}^{15}\sum_{j=1}^{378}  b_{ij} \hspace{0.2pt} R^{(k=1)}_i \hspace{0.5pt} \cI^{(2)}_j, \notag\\
\mathcal{A}_{7;\hspace{0.2pt}2}^{(2)}&=\sum_{i=1}^{15}\sum_{j=1}^{378}  c_{ij}\hspace{0.2pt}  R^{(k=2)}_i \hspace{0.5pt} \cI^{(2)}_j,
\qquad& \mathcal{A}_{7;\hspace{0.2pt}3}^{(2)}&=\mathcal{A}_{7;\hspace{0.2pt}3}^{(0)}\sum_{j=1}^{378} d_j \hspace{0.5pt} \cI^{(2)}_j , \vspace{-0.5pt} \label{twoloop7ptans}
\end{align}
with $(1\hspace{-1pt}+\hspace{-1pt}15\hspace{-1pt}+\hspace{-1pt}15\hspace{-1pt}+\hspace{-1pt}1)\hspace{-1pt}\times\hspace{-1pt} 378\!=\!12,096$ free coefficients, together with the one-loop result (with its 6 free coefficients) into the duality equation, which at this loop order reads:
\eq{\lim_{\substack{\text{7-gon}\\\text{light-like}}}\!\! \xi^{(7)}\mathcal{F}^{(5)}=\cAA{7}{0}^{(2)}+\frac{\cAA{7}{3}^{(2)}}{\cAA{7}{3}^{(0)}} +\frac{\cAA{7}{0}^{(1)}\,\cAA{7}{3}^{(1)}}{\cAA{7}{3}^{(0)}} +\frac{\cAA{7}{1}^{(2)}\,\cAA{7}{2}^{(0)}}{\cAA{7}{3}^{(0)}}+\frac{\cAA{7}{2}^{(2)}\,\cAA{7}{1}^{(0)}}{\cAA{7}{3}^{(0)}}+\frac{\cAA{7}{1}^{(1)}\,\cAA{7}{2}^{(1)}}{\cAA{7}{3}^{(0)}}.\label{seven_point_lightlike_correlator_five_loops}}
The solution has  $378\!+ \!378 \!= \!756$ free coefficients, $378$ parameters for NMHV/N$^2$MHV and $378$ for MHV/N$^3$MHV consistent with the ambiguity~\eqref{amb}. Imposing parity invariance reduces this to  $168 \!+ \!168\! =\!  336 $ free coefficients, with $168\!=\!7\!\times\! 24$ being the number of independent parity-odd integrands in the prescriptive basis. Finally, imposing cyclic invariance in addition yields a final solution with  $24 \!+ \!24 \!=\! 48 $ parameters at two loops, with 24 understood as the number of cyclic families of parity-odd integrands. In the process of doing so, the remaining one-loop sector is obtained (up to a sign ambiguity of  the parity-odd integrands of the (anti-)MHV amplitudes as seen at six points~\eqref{MHV1_MHVbar1}.)
 We expect these 48 remaining coefficients to be fixed by going one loop higher. 
 We reiterate that at this order, the correlator/amplitude duality~\eqref{seven_point_lightlike_correlator_five_loops} solves the seven-point parity-even part of the amplitude up to two loops.
 
\vspace{-6pt}\section{Conclusions}\label{sec:outlook}\vspace{-4pt}

One consequence of the correlator/amplitude duality is that the simplest (four-point) correlator contains a certain combination of all $n$-point superamplitudes for any $n$. 
In this paper, we provide evidence for the conjecture that this combination contains all the information from the individual amplitudes---the four-point correlator contains all information about every amplitude integrand. We show this by extracting the individual amplitudes from the null correlator. From the correlator to four loops we extract the six particle tree-level, one-loop and parity-even part of the two-loop amplitude. From the correlator up to five loops we extract the six and seven particle tree-level, one-loop and parity-even part of the two-loop amplitude. An obvious future direction is to test this at higher loops/points.

To perform the extraction of individual amplitudes at six and seven points, we compared to an ansatz for the amplitudes and resorted to numerical evaluation of the rational integrands and solved the resulting equations.
This method is in stark contrast to the extraction of  four~\cite{1201.5329} and five~\cite{1312.1163} point amplitudes from the correlator,  where the duality is seen 
algebraically (rather than just numerically), and in the four-point and five-point parity-odd case, even graphically. In these cases there are simple graphical rules for determining all amplitude integrand graphs from the correlator $f$ graphs without ever needing to introduce an ansatz. Consistency of these graphical amplitude extraction rules with the hidden symmetry inherent in the $f$-graph structure led to the discovery of graphical rules which gave  the higher-loop correlator in terms of the lower-loop one~\cite{1609.00007}. The (vastly efficient) graphical nature of these procedures allows for the determination of the four-point correlator to ten loops.

The next step left for future work is to attempt to understand the higher-point duality discussed here from a more algebraic or even graphical perspective. The main complication is the presence of spurious poles in the basis of Yangian invariants that appear from NMHV and onwards. These \textit{must} cancel in the sum, but this is difficult to see algebraically and requires non-trivial algebraic identities, thus spoiling a transparent approach. Nevertheless,  it may still be  possible to read off graphically, directly from the correlator, certain integrands (with their coefficients) which appear in the amplitudes.

Another complication that appears from six points is that it is no longer automatically clear from the topology of a graph whether it contributes to a particular loop amplitude or to the product of lower-loop amplitudes.

\begin{figure}[h!]
  \centering
  \begin{minipage}[b]{0.18\textwidth}
    \includegraphics[width=\textwidth]{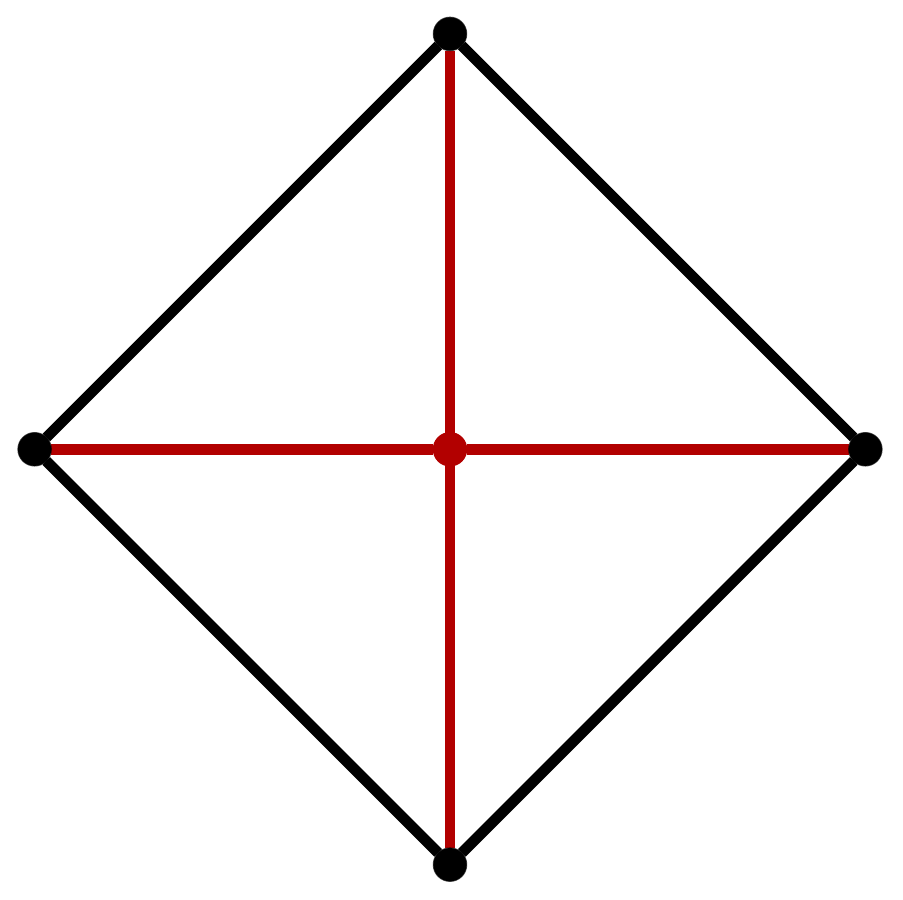}
  \end{minipage}
    \hspace{0.75cm}
  \begin{minipage}[b]{0.18\textwidth}
    \includegraphics[width=\textwidth]{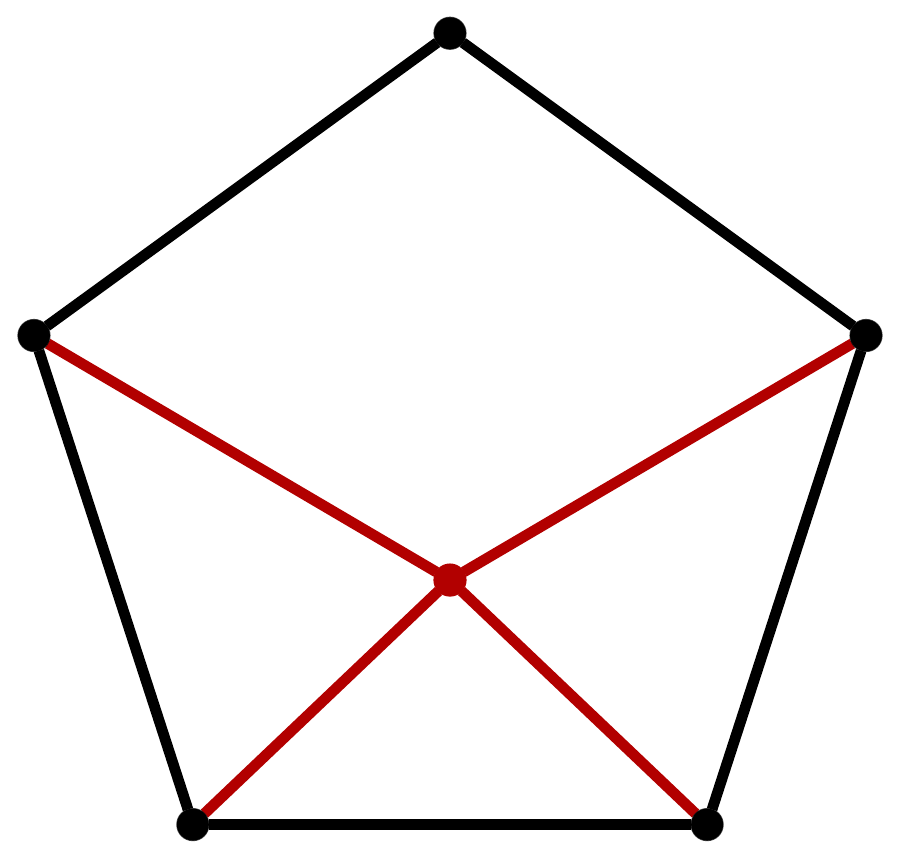}
  \end{minipage}
     \hspace{0.8cm}
  \begin{minipage}[b]{0.18\textwidth}
    \includegraphics[width=\textwidth]{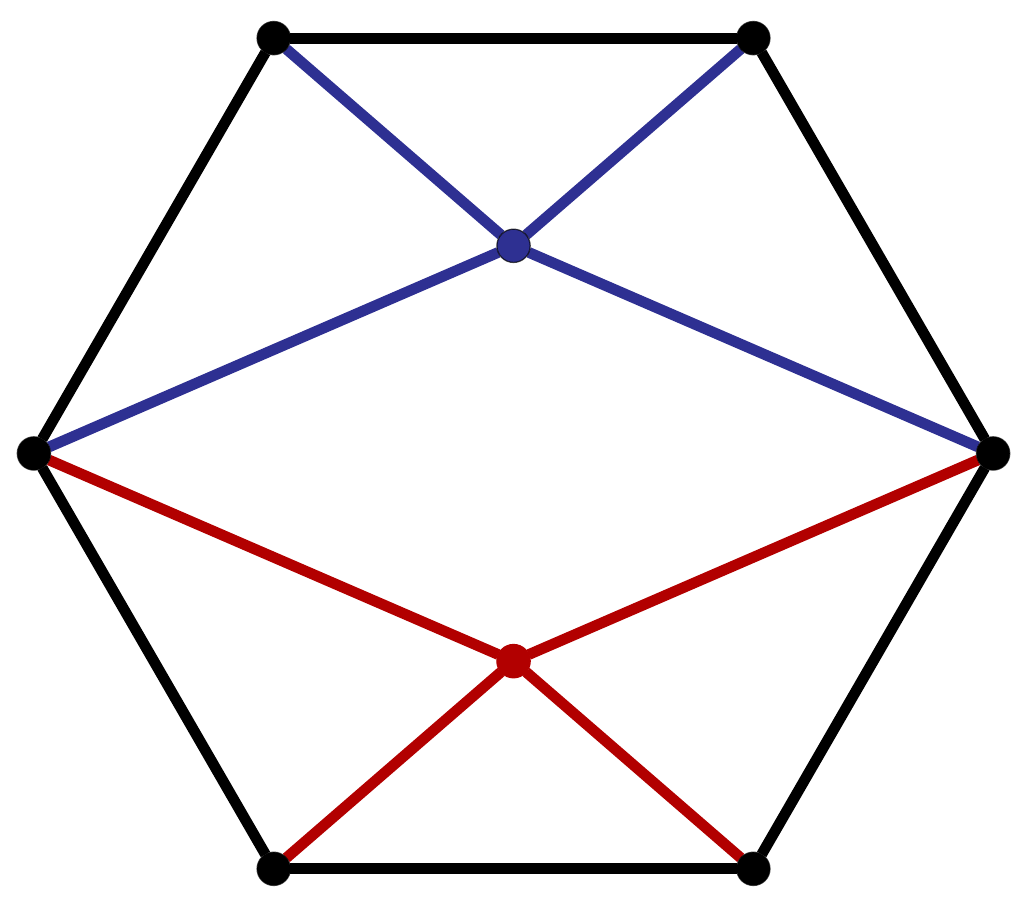}
   \end{minipage}
      \hspace{0.8cm}
     \begin{minipage}[b]{0.18\textwidth}
    \includegraphics[width=\textwidth]{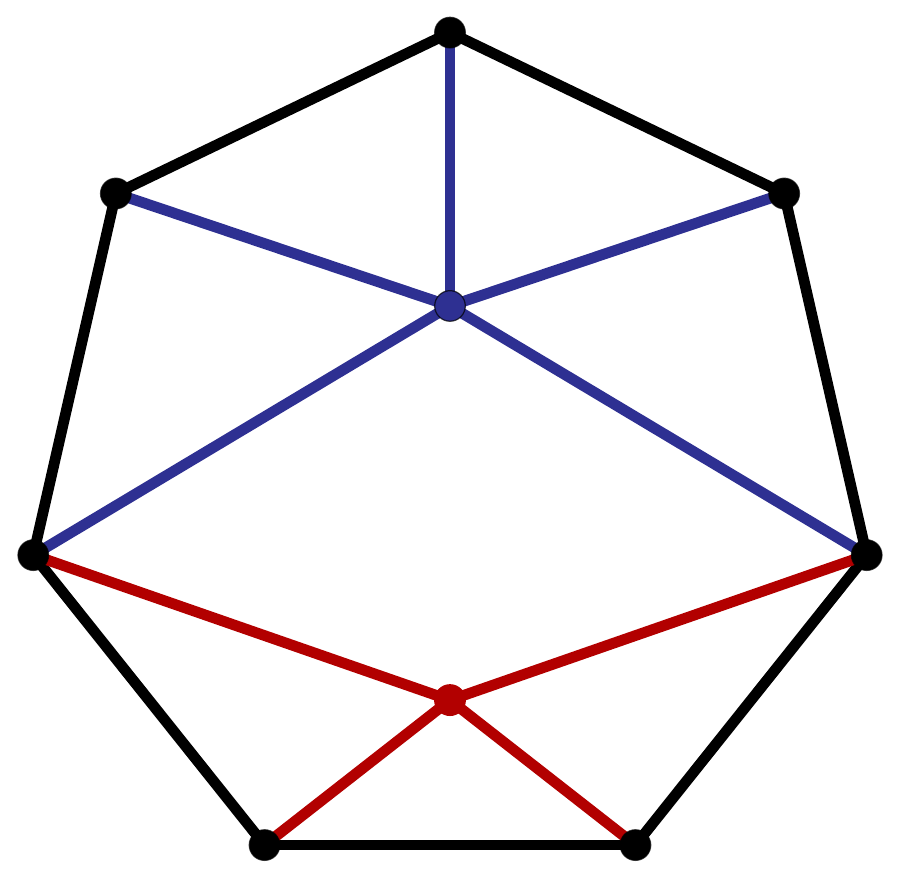}
   \end{minipage}
   \captionsetup{width=0.8\linewidth}
   \caption{Disc planar contributions of highlighted-box integrands to four, five, six and seven (light-like) cycles.} \label{fig:four_five_six_seven_cycle_box}
  \end{figure}
For example, in \mbox{Figure \ref{fig:four_five_six_seven_cycle_box}}, the third graph in the figure could arise from a one-loop times one-loop product, or be a  two-loop graph contribution. At four- and five-points, planarity ruled out such ambiguities.

With or  without such a graphical approach however, this four-point correlator approach to obtaining amplitudes provides a highly efficient method for obtaining amplitudes directly as local integrands at any number of points as well as informing us something highly non-trivial about their structure. Note that the four-point correlator can be obtained directly from the four-point amplitude, and so one can phrase this as a method for obtaining all amplitudes from the four-point amplitude!

Finally, it would be very interesting to combine these considerations with the geometrical polyhedral approach of the  amplituhedron~\cite{1312.2007,1312.7878}. In particular, all the squared amplitudes in~\cite{1701.00453} were found to be given by explicit geometrical regions in an appropriate Grassmannian. Combining this with additional topological information---the winding number of~\cite{1704.05069}---may give the separate contributions in the sum~\eqref{n_point_dualitya}.

\vspace{-0pt}\section*{Acknowledgements}\vspace{-6pt}
 The authors gratefully acknowledge many helpful discussions  with Jacob Bourjaily. We also thank Tomasz Lukowski for helpful suggestions for obtaining  Yangian invariants. This work was supported by an STFC studentship and a GATIS visitors fund (VVT); and by an STFC Consolidated Grant ST/L000407/1 and the Marie Curie network GATIS (gatis.desy.eu) of the European Union's Seventh Framework Programme FP7/2007-2013 under REA Grant Agreement No.\ 317089 (PH). VVT would also like to acknowledge the hospitality of Institute Physical Th\'eorique, UMR 3681, and Laboratoire de Physique  Th\'eorique LPTENS, UMR 8549, where this work began, particularly highlighting the kindness shown by Benjamin Basso, Ivan Kostov, and Didina Serban.

\appendix

%
\vspace{-0pt}\section{Seven-Point N${}^2$MHV Yangian invariants} \label{appendix_covariantisation}\vspace{-3pt}
We use the following co-ordinates for $\hat Z$ (using $\GL(6)$ invariance) throughout this appendix
\eq{\hat Z_{a}^{\tilde{A}} =\begin{bmatrix}1 & 0 & 0 & 0 & 0 & 0 & A \\[-0.5ex]
0 & 1 & 0 & 0 & 0 & 0 & B \\[-0.5ex]
0 & 0 & 1 & 0 & 0 & 0 & C \\[-0.5ex]
0 & 0 & 0 & 1 & 0 & 0 & D \\[-0.5ex]
0 & 0 & 0 & 0 & 1 & 0 & E \\[-0.5ex]
0 & 0 & 0 & 0 & 0 & 1 & F 
\end{bmatrix},}
so that
\eq{\begin{gathered}
A=-\sixBra{2}{3}{4}{5}{6}{7},\hspace{5pt}B=\sixBra{1}{3}{4}{5}{6}{7},\hspace{5pt}C=-\sixBra{1}{2}{4}{5}{6}{7},\\
D=\sixBra{1}{2}{3}{5}{6}{7},\hspace{5pt}E=-\sixBra{1}{2}{3}{4}{6}{7},\hspace{5pt}F=\sixBra{1}{2}{3}{4}{5}{7},\hspace{5pt} 1= \sixBra{1}{2}{3}{4}{5}{6}.
\end{gathered} \label{grassmannian_capital_relations_1234}}
%
\vspace{-2pt}\subsection{$(12)=(34)=0$ Residue} \label{appendix_B1}
\textit{Canonical} positive co-ordinates on the Grassmannian are \cite{1212.5605}
\eq{\begin{aligned}
C_{\alpha a}&=\begin{bmatrix} 1 & \alpha_8\hspace{-1.5pt} & \alpha_2 \hspace{-1.5pt}+\hspace{-1.5pt}\alpha_4 \hspace{-1.5pt}+\hspace{-1.5pt}\alpha_6 & (\alpha_2\hspace{-1.5pt}+\hspace{-1.5pt}\alpha_4 + \hspace{-1.5pt}\alpha_6)\alpha_7 &(\alpha_2 + \hspace{-1.5pt}\alpha_4)\alpha_5 & \alpha_2 \alpha_3 & 0 \\
0 & 0 & 1 & \alpha_7 & \alpha_5 & \alpha_3 & \alpha_1 
\end{bmatrix} \\
\Rightarrow Y_{\alpha}^{\tilde{A}}&=\begin{bmatrix} 1 & \alpha_8 & \alpha_2\hspace{-1.5pt}+\hspace{-1.5pt}\alpha_4\hspace{-1.5pt}+\hspace{-1.5pt}\alpha_6 & (\alpha_2\hspace{-1.5pt}+\hspace{-1.5pt}\alpha_4\hspace{-1.5pt} +\hspace{-1.5pt}\alpha_6)\alpha_7 & (\alpha_2\hspace{-1.5pt}+\hspace{-1.5pt}\alpha_4)\alpha_5  & \alpha_2\alpha_3   \\
A \alpha_1 & B \alpha_1 & 1\hspace{-1.5pt}+\hspace{-1.5pt}C\alpha_1 & D\alpha_1\hspace{-1.5pt}+\hspace{-1.5pt}\alpha_7 & E\alpha_1 \hspace{-1.5pt}+\hspace{-1.5pt}\alpha_5 & F\alpha_1 \hspace{-1.5pt}+\hspace{-1.5pt}\alpha_3 
\end{bmatrix}. \end{aligned}}
The $Y$-weighted differential form is found to be
\eq{\twoBra{Y}{d^4 Y_1}\twoBra{Y}{d^4 Y_2} = \alpha_1 \alpha_3 \alpha_5 (D\hspace{-1.5pt}-\hspace{-1.5pt}C\alpha_7)(B\hspace{-1.5pt}-\hspace{-1.5pt}A\alpha_8)\hspace{1.5pt} \text{d}\alpha_1 \dots \text{d}\alpha_8.  \label{Y_form_1234}} 
The differential form is normalised to be $Y$-weightless
\eq{\frac{\twoBra{Y}{d^4 Y_1}\twoBra{Y}{d^4 Y_2}}{\fiveBra{Y}{2}{5}{6}{7} \fiveBra{Y}{3}{4}{5}{6}  \fiveBra{Y}{3}{4}{5}{7} \fiveBra{Y}{3}{4}{6}{7} \fiveBra{Y}{3}{5}{6}{7} \fiveBra{Y}{4}{5}{6}{7}} =  -\frac{\text{d}\alpha_1 \dots \text{d}\alpha_8}{\alpha_7 (B - A \alpha_8)^4}.  \label{Y_form_weightless_1234}} 
Therefore, the $(12)=(34)=0$ residue is given as
\eq{\Omega_{(12),(34)}=-\frac{(B-A \alpha_8)^4\twoBra{Y}{d^4 Y_1}\twoBra{Y}{d^4 Y_2}}{\alpha_1 \alpha_2 \alpha_3 \alpha_4 \alpha_5 \alpha_6 \alpha_8 \hspace{1.5pt} \fiveBra{Y}{2}{5}{6}{7} \fiveBra{Y}{3}{4}{5}{6}  \fiveBra{Y}{3}{4}{5}{7} \fiveBra{Y}{3}{4}{6}{7} \fiveBra{Y}{3}{5}{6}{7} \fiveBra{Y}{4}{5}{6}{7}} .  \label{amplituhedron_1234_first}} 
The $Y$-weightless cross ratios for the positive co-ordinates are:
\eq{\begin{aligned}
&\alpha_1 = \frac{\fiveBra{Y}{3}{4}{5}{6}}{\fiveBra{Y}{4}{5}{6}{7}},\hspace{5pt}\alpha_2= \frac{\fiveBra{Y}{4}{5}{6}{7}\left(\fiveBra{Y}{1}{2}{5}{7}\fiveBra{Y}{3}{4}{5}{6}\hspace{-1.5pt}-\hspace{-1.5pt}\fiveBra{Y}{1}{2}{5}{6}\fiveBra{Y}{3}{4}{5}{7}\right)}{\fiveBra{Y}{2}{5}{6}{7}\fiveBra{Y}{3}{4}{5}{6}\fiveBra{Y}{3}{4}{5}{7}},\hspace{5pt}\\\ &\alpha_3 = \frac{\fiveBra{Y}{3}{4}{7}{5}}{\fiveBra{Y}{4}{5}{6}{7}}, \hspace{5pt}\alpha_4= \frac{\fiveBra{Y}{4}{5}{6}{7}\left(\fiveBra{Y}{1}{2}{6}{7}\fiveBra{Y}{3}{4}{5}{7}\hspace{-1.5pt}-\hspace{-1.5pt}\fiveBra{Y}{1}{2}{5}{7}\fiveBra{Y}{3}{4}{6}{7}\right)}{\fiveBra{Y}{2}{5}{6}{7}\fiveBra{Y}{3}{4}{5}{7}\fiveBra{Y}{3}{4}{6}{7}},\\
&\alpha_5 = \frac{\fiveBra{Y}{3}{4}{6}{7}}{\fiveBra{Y}{4}{5}{6}{7}},\hspace{5pt}\alpha_6= -\frac{\fiveBra{Y}{1}{2}{6}{7}\fiveBra{Y}{4}{5}{6}{7}}{\fiveBra{Y}{2}{5}{6}{7}\fiveBra{Y}{3}{4}{6}{7}},\hspace{5pt}\alpha_7= -\frac{\fiveBra{Y}{3}{5}{6}{7}}{\fiveBra{Y}{4}{5}{6}{7}},\hspace{5pt}\alpha_8= -\frac{\fiveBra{Y}{1}{5}{6}{7}}{\fiveBra{Y}{2}{5}{6}{7}}.
\end{aligned} \label{grassmannian_coord_relations_1234}}
Substituting these into (\ref{amplituhedron_1234_first}) yields a covariant expression for the residue
\eq{\frac{ \twoBra{Y}{d^4 Y_1}\twoBra{Y}{d^4 Y_2}( \langle Y \hspace{0.5pt}[ 1 | \hspace{0.5pt} 5 \hspace{0.5pt}  6  \hspace{0.5pt} 7 \rangle \langle | 2 ] \hspace{0.5pt} 3 \hspace{0.5pt} 4 \hspace{0.5pt}5 \hspace{0.5pt} 6 \hspace{0.5pt} 7 \rangle)^4}{\fiveBra{Y}{1}{2}{6}{7}\fiveBra{Y}{1}{5}{6}{7}\fiveBra{Y}{2}{5}{6}{7}\fiveBra{Y}{3}{4}{5}{6}\fiveBra{Y}{3}{5}{6}{7}\fiveBra{Y}{4}{5}{6}{7}
\langle Y \hspace{0.5pt}  1 \hspace{0.5pt} 2 \hspace{0.5pt} 5 [ 7 | \rangle \langle Y \hspace{0.5pt} 3 \hspace{0.5pt} 4 \hspace{0.5pt} 5 | 6 ] \rangle \langle Y \hspace{0.5pt}  1 \hspace{0.5pt} 2  \hspace{0.5pt} [ 6 | 7  \rangle \langle Y \hspace{0.5pt} 3 \hspace{0.5pt} 4  | 5 ] \hspace{0.5pt} 7 \rangle}. \label{amplituhedron_1234_final} \notag} 
%
\vspace{-2pt}\subsection{$(12)=(45)=0$ Residue}\label{appendix_B2}

\textit{Canonical} positive co-ordinates on the Grassmannian are \cite{1212.5605}
\eq{\begin{aligned}
C_{\alpha a}&=\begin{bmatrix} 1 & \alpha_8\hspace{-1.5pt} & \alpha_2 \hspace{-1.5pt}+\hspace{-1.5pt}\alpha_4 \hspace{-1.5pt}+\hspace{-1.5pt}\alpha_7 & (\alpha_2\hspace{-1.5pt}+\hspace{-1.5pt}\alpha_4)\alpha_6 &(\alpha_2 + \hspace{-1.5pt}\alpha_4)\alpha_5 & \alpha_2 \alpha_3 & 0 \\
0 & 0 & 1 & \alpha_6 & \alpha_5 & \alpha_3 & \alpha_1 
\end{bmatrix} \\
\Rightarrow Y_{\alpha}^{\tilde{A}}&=\begin{bmatrix} 1 & \alpha_8 & \alpha_2\hspace{-1.5pt}+\hspace{-1.5pt}\alpha_4\hspace{-1.5pt}+\hspace{-1.5pt}\alpha_7 & (\alpha_2\hspace{-1.5pt}+\hspace{-1.5pt}\alpha_4)\alpha_6 & (\alpha_2\hspace{-1.5pt}+\hspace{-1.5pt}\alpha_4)\alpha_5  & \alpha_2\alpha_3   \\
A \alpha_1 & B \alpha_1 & 1\hspace{-1.5pt}+\hspace{-1.5pt}C\alpha_1 & D\alpha_1\hspace{-1.5pt}+\hspace{-1.5pt}\alpha_6 & E\alpha_1 \hspace{-1.5pt}+\hspace{-1.5pt}\alpha_5 & F\alpha_1 \hspace{-1.5pt}+\hspace{-1.5pt}\alpha_3 
\end{bmatrix}. \end{aligned}}
The $Y$-weighted differential form is found to be
\eq{\twoBra{Y}{d^4 Y_1}\twoBra{Y}{d^4 Y_2} = \alpha_1 \alpha_3 (D\alpha_5 \hspace{-1.5pt}-\hspace{-1.5pt}E\alpha_6)(B\hspace{-1.5pt}-\hspace{-1.5pt}A\alpha_8)\hspace{1.5pt} \text{d}\alpha_1 \dots \text{d}\alpha_8.  \label{Y_form_1245}} 
The differential form is normalised to be $Y$-weightless
\eq{\frac{\twoBra{Y}{d^4 Y_1}\twoBra{Y}{d^4 Y_2}}{\fiveBra{Y}{1}{2}{6}{7} \fiveBra{Y}{3}{4}{5}{6}  \fiveBra{Y}{3}{4}{5}{7} \fiveBra{Y}{3}{4}{6}{7} \fiveBra{Y}{3}{5}{6}{7} \fiveBra{Y}{4}{5}{6}{7}} =  \frac{\text{d}\alpha_1 \dots \text{d}\alpha_8}{\alpha_5 \alpha_6 \alpha_7 (B - A \alpha_8)^4}.  \label{Y_form_weightless_1245}} 
Therefore, the $(12)=(45)=0$ residue is given as
\eq{\Omega_{(12),(45)}=\frac{(B-A \alpha_8)^4\twoBra{Y}{d^4 Y_1}\twoBra{Y}{d^4 Y_2}}{\alpha_1 \alpha_2 \alpha_3 \alpha_4 \alpha_8 \hspace{1.5pt} \fiveBra{Y}{1}{2}{6}{7} \fiveBra{Y}{3}{4}{5}{6}  \fiveBra{Y}{3}{4}{5}{7} \fiveBra{Y}{3}{4}{6}{7} \fiveBra{Y}{3}{5}{6}{7} \fiveBra{Y}{4}{5}{6}{7}} .  \label{amplituhedron_1245_first}} 
The cross ratios for the positive co-ordinates are:
\eq{\begin{aligned}
&\alpha_1 = \frac{\fiveBra{Y}{3}{4}{5}{6}}{\fiveBra{Y}{4}{5}{6}{7}},\hspace{5pt}\alpha_2= \frac{\fiveBra{Y}{4}{5}{6}{7}\left(\fiveBra{Y}{1}{2}{3}{7}\fiveBra{Y}{3}{4}{5}{6}\hspace{-1.5pt}-\hspace{-1.5pt}\fiveBra{Y}{1}{2}{3}{6}\fiveBra{Y}{3}{4}{5}{7}\right)}{\fiveBra{Y}{2}{3}{6}{7}\fiveBra{Y}{3}{4}{5}{6}\fiveBra{Y}{3}{4}{5}{7}},\hspace{5pt}\\\ &\alpha_3 = \frac{\fiveBra{Y}{3}{4}{7}{5}}{\fiveBra{Y}{4}{5}{6}{7}}, \hspace{5pt}\alpha_4=- \frac{\fiveBra{Y}{1}{2}{3}{7}\fiveBra{Y}{4}{5}{6}{7}}{\fiveBra{Y}{2}{3}{6}{7}\fiveBra{Y}{3}{4}{5}{7}},\hspace{5pt}\alpha_5 = \frac{\fiveBra{Y}{3}{4}{6}{7}}{\fiveBra{Y}{4}{5}{6}{7}},\hspace{5pt}\alpha_6= -\frac{\fiveBra{Y}{3}{5}{6}{7}}{\fiveBra{Y}{4}{5}{6}{7}},\\ &\alpha_7= \frac{\fiveBra{Y}{1}{2}{6}{7}\fiveBra{Y}{4}{5}{6}{7}}{\sixBra{1}{2}{3}{5}{6}{7}\fiveBra{Y}{3}{4}{6}{7}-\sixBra{1}{2}{3}{4}{6}{7}\fiveBra{Y}{3}{5}{6}{7}},\hspace{5pt}\alpha_8= -\frac{\fiveBra{Y}{1}{3}{6}{7}}{\fiveBra{Y}{2}{3}{6}{7}}.
\end{aligned} \label{grassmannian_coord_relations_1245}}
We note that $\alpha_7$ is not weightless in $Y$ but (\ref{amplituhedron_1245_first}) is independent of $\alpha_7$. Substituting these into (\ref{amplituhedron_1245_first}) yields the following
\eq{\frac{ \twoBra{Y}{d^4 Y_1}\twoBra{Y}{d^4 Y_2}( \langle Y \hspace{0.5pt}[ 2 | \hspace{0.5pt} 3 \hspace{0.5pt}  6  \hspace{0.5pt} 7 \rangle \langle | 1 ] \hspace{0.5pt} 3 \hspace{0.5pt} 4 \hspace{0.5pt}5 \hspace{0.5pt} 6 \hspace{0.5pt} 7 \rangle)^4}{\fiveBra{Y}{1}{2}{3}{7}\fiveBra{Y}{1}{2}{6}{7}\fiveBra{Y}{1}{3}{6}{7}\fiveBra{Y}{2}{3}{6}{7}\fiveBra{Y}{3}{4}{5}{6}\fiveBra{Y}{3}{4}{6}{7}\fiveBra{Y}{3}{5}{6}{7}\fiveBra{Y}{4}{5}{6}{7}
\langle Y \hspace{0.5pt}  1 \hspace{0.5pt} 2 \hspace{0.5pt} 3 [ 7 | \rangle \langle Y \hspace{0.5pt} 3 \hspace{0.5pt} 4 \hspace{0.5pt} 5 | 6 ] \rangle} . \notag \label{amplituhedron_1245_final}} 
%

\providecommand{\href}[2]{#2}\begingroup\raggedright\endgroup


\begin{thebibliography}{10}

\bibitem{hep-th/9811155}
F.~Gonzalez-Rey, I.~Y. Park, and K.~Schalm, ``{A Note on Four Point Functions
  of Conformal Operators in $\mathcal{N}\!=\!4$ Super Yang-Mills},''
  \href{http://dx.doi.org/10.1016/S0370-2693(99)00017-9}{{\em Phys. Lett.} {\bf
  B448} (1999)  37--40},
\href{http://arxiv.org/abs/hep-th/9811155}{{ arXiv:hep-th/9811155 [hep-th]}}.

\bibitem{hep-th/9811172}
B.~Eden, P.~S. Howe, C.~Schubert, E.~Sokatchev, and P.~C. West, ``{Four Point
  Functions in $\mathcal{N}\!=\!4$ Supersymmetric Yang-Mills Theory at Two
  Loops},'' \href{http://dx.doi.org/10.1016/S0550-3213(99)00360-0}{{\em Nucl.
  Phys.} {\bf B557} (1999)  355--379},
\href{http://arxiv.org/abs/hep-th/9811172}{{ arXiv:hep-th/9811172 [hep-th]}}.

\bibitem{hep-th/9906051}
B.~Eden, P.~S. Howe, C.~Schubert, E.~Sokatchev, and P.~C. West,
  ``{Simplifications of Four Point Functions in $\mathcal{N}\!=\!4$
  Supersymmetric Yang-Mills Theory at Two Loops},''
  \href{http://dx.doi.org/10.1016/S0370-2693(99)01033-3}{{\em Phys. Lett.} {\bf
  B466} (1999)  20--26},
\href{http://arxiv.org/abs/hep-th/9906051}{{ arXiv:hep-th/9906051 [hep-th]}}.

\bibitem{hep-th/0003096}
B.~Eden, C.~Schubert, and E.~Sokatchev, ``{Three Loop four Point Correlator in
  $\mathcal{N}\!=\!4$ SYM},''
  \href{http://dx.doi.org/10.1016/S0370-2693(00)00515-3}{{\em Phys. Lett.} {\bf
  B482} (2000)  309--314},
\href{http://arxiv.org/abs/hep-th/0003096}{{ arXiv:hep-th/0003096 [hep-th]}}.

\bibitem{1108.3557}
B.~Eden, P.~Heslop, G.~P. Korchemsky, and E.~Sokatchev, ``{Hidden Symmetry of
  Four-Point Correlation Functions and Amplitudes in $\mathcal{N}\!=\!4$
  SYM},'' \href{http://dx.doi.org/10.1016/j.nuclphysb.2012.04.007}{{\em Nucl.
  Phys.} {\bf B862} (2012)  193--231},
\href{http://arxiv.org/abs/1108.3557}{{ arXiv:1108.3557 [hep-th]}}.

\bibitem{1201.5329}
B.~Eden, P.~Heslop, G.~P. Korchemsky, and E.~Sokatchev, ``{Constructing the
  Correlation Function of Four Stress-Tensor Multiplets and the Four-Particle
  Amplitude in $\mathcal{N}\!=\!4$ SYM},''
  \href{http://dx.doi.org/10.1016/j.nuclphysb.2012.04.013}{{\em Nucl. Phys.}
  {\bf B862} (2012)  450--503},
\href{http://arxiv.org/abs/1201.5329}{{ arXiv:1201.5329 [hep-th]}}.

\bibitem{1303.6909}
J.~Drummond, C.~Duhr, B.~Eden, P.~Heslop, J.~Pennington, and V.~A. Smirnov,
  ``{Leading Singularities and Off-Shell Conformal Integrals},''
  \href{http://dx.doi.org/10.1007/JHEP08(2013)133}{{\em JHEP} {\bf 08} (2013)
  133},
\href{http://arxiv.org/abs/1303.6909}{{ arXiv:1303.6909 [hep-th]}}.

\bibitem{1512.07912}
J.~L. Bourjaily, P.~Heslop, and V.-V. Tran, ``{Perturbation Theory at Eight
  Loops: Novel Structures and the Breakdown of Manifest Conformality in
  $\mathcal{N}\!=\!4$ Supersymmetric Yang-Mills Theory},''
  \href{http://dx.doi.org/10.1103/PhysRevLett.116.191602}{{\em Phys. Rev.
  Lett.} {\bf 116} (2016) no. 19, 191602},
\href{http://arxiv.org/abs/1512.07912}{{ arXiv:1512.07912 [hep-th]}}.

\bibitem{1609.00007}
J.~L. Bourjaily, P.~Heslop, and V.-V. Tran, ``{Amplitudes and Correlators to Ten Loops Using Simple, Graphical Bootstraps},''
  \href{http://dx.doi.org/10.1007/JHEP11(2016)125}{{\em JHEP} {\bf 1611} (2016) 125},
\href{http://arxiv.org/abs/1609.00007}{{ arXiv:1609.00007 [hep-th]}}.

\bibitem{1303.4734}
J.~L. Bourjaily, S. ~Caron-Huot, and J.~Trnka
  ``{Dual-Conformal Regularization of Infrared Loop Divergences and the Chiral Box Expansion},''
  \href{https://doi.org/10.1007/JHEP01(2015)001}{{\em JHEP} {\bf 1501} (2015)
   001},
\href{https://arxiv.org/abs/1303.4734}{{ arXiv:1303.4734 [hep-th]}}.

\bibitem{1505.05886}
J.~L. Bourjaily and J.~Trnka
  ``{Local Integrand Representations of All Two-Loop Amplitudes in Planar SYM},''
  \href{https://doi.org/10.1007/JHEP08(2015)119}{{\em JHEP} {\bf 08} (2015)
   119},
\href{https://arxiv.org/abs/1505.05886}{{ arXiv:1505.05886 [hep-th]}}.

\bibitem{1704.05460}
J.~L. Bourjaily, E.~Herrmann, and J.~Trnka
  ``{Prescriptive Unitarity},''
  \href{https://doi.org/10.1007/JHEP06(2017)059}{{\em JHEP} {\bf 06} (2017)
   059},
\href{https://arxiv.org/abs/1704.05460}{{ arXiv:1704.05460 [hep-th]}}.

\bibitem{1012.6032}
N.~Arkani-Hamed, J.~L. Bourjaily, F.~Cachazo, and J.~Trnka, ``{Local Integrals
  for Planar Scattering Amplitudes},''
  \href{http://dx.doi.org/10.1007/JHEP06(2012)125}{{\em JHEP} {\bf 1206} (2012)
   125},
\href{http://arxiv.org/abs/1012.6032}{{ arXiv:1012.6032 [hep-th]}}.

\bibitem{1008.2958}
N.~Arkani-Hamed, J.~L. Bourjaily, F.~Cachazo, S.~Caron-Huot, and J.~Trnka,
  ``{The All-Loop Integrand For Scattering Amplitudes in Planar
  $\mathcal{N}\!=\!4$ SYM},''
  \href{http://dx.doi.org/10.1007/JHEP01(2011)041}{{\em JHEP} {\bf 1101} (2011)
   041},
\href{http://arxiv.org/abs/1008.2958}{{ arXiv:1008.2958 [hep-th]}}.

\bibitem{ArkaniHamed:book}
N.~Arkani-Hamed, J.~L. Bourjaily, F.~Cachazo, A.~B. Goncharov, A.~Postnikov,
  and J.~Trnka, {\em {Grassmannian Geometry of Scattering Amplitudes}}.
\newblock Cambridge University Press, 2016.

\bibitem{1312.1163}
R.~G. Ambrosio, B.~Eden, T.~Goddard, P.~Heslop, and C.~Taylor, ``{Local
  Integrands for the Five-Point Amplitude in Planar $\mathcal{N}\!=\!4$ SYM Up
  to Five Loops},'' \href{http://dx.doi.org/10.1007/JHEP01(2015)116}{{\em JHEP}
  {\bf 01} (2015)  116},
\href{http://arxiv.org/abs/1312.1163}{{ arXiv:1312.1163 [hep-th]}}.

\bibitem{1007.3243}
L.~F. Alday, B.~Eden, G.~P. Korchemsky, J.~Maldacena, and E.~Sokatchev, ``{From
  Correlation Functions to Wilson Loops},''
  \href{http://dx.doi.org/10.1007/JHEP09(2011)123}{{\em JHEP} {\bf 1109} (2011)
   123},
\href{http://arxiv.org/abs/1007.3243}{{ arXiv:1007.3243 [hep-th]}}.

\bibitem{1007.3246}
B.~Eden, G.~P. Korchemsky, and E.~Sokatchev, ``{From Correlation Functions to
  Scattering Amplitudes},''
  \href{http://dx.doi.org/10.1007/JHEP12(2011)002}{{\em JHEP} {\bf 1112} (2011)
   002},
\href{http://arxiv.org/abs/1007.3246}{{ arXiv:1007.3246 [hep-th]}}.

\bibitem{1009.2488}
B.~Eden, G.~P. Korchemsky, and E.~Sokatchev, ``{More on the Duality
  Correlators/Amplitudes},''
  \href{http://dx.doi.org/10.1016/j.physletb.2012.02.014}{{\em Phys. Lett.}
  {\bf B709} (2012)  247--253},
\href{http://arxiv.org/abs/1009.2488}{{ arXiv:1009.2488 [hep-th]}}.

\bibitem{1103.3714}
B.~Eden, P.~Heslop, G.~P. Korchemsky, and E.~Sokatchev, ``{The
  Super-Correlator/ Super-Amplitude Duality: Part I},''
  \href{http://dx.doi.org/10.1016/j.nuclphysb.2012.12.015}{{\em Nucl. Phys.}
  {\bf B869} (2013)  329--377},
\href{http://arxiv.org/abs/1103.3714}{{ arXiv:1103.3714 [hep-th]}}.

\bibitem{1103.4119}
T.~Adamo, M.~Bullimore, L.~Mason, and D.~Skinner, ``{A Proof of the
  Supersymmetric Correlation Function / Wilson Loop Correspondence},''
  \href{http://dx.doi.org/10.1007/JHEP08(2011)076}{{\em JHEP} {\bf 1108} (2011)
   076},
\href{http://arxiv.org/abs/1103.4119}{{ arXiv:1103.4119 [hep-th]}}.

\bibitem{1103.4353}
B.~Eden, P.~Heslop, G.~P. Korchemsky, and E.~Sokatchev, ``{The
  Super-Correlator/ Super-Amplitude Duality: Part II},''
  \href{http://dx.doi.org/10.1016/j.nuclphysb.2012.12.014}{{\em Nucl. Phys.}
  {\bf B869} (2013)  378--416},
\href{http://arxiv.org/abs/1103.4353}{{ arXiv:1103.4353 [hep-th]}}.

\bibitem{0909.0250}
L.~Mason and D.~Skinner, ``{Dual Superconformal Invariance, Momentum Twistors
  and Grassmannians},''
  \href{http://dx.doi.org/10.1088/1126-6708/2009/11/045}{{\em JHEP} {\bf 0911}
  (2009)  045},
\href{http://arxiv.org/abs/0909.0250}{{ arXiv:0909.0250 [hep-th]}}.

\bibitem{0912.3249}
N.~Arkani-Hamed, J.~Bourjaily, F.~Cachazo, and J.~Trnka, ``{Local Spacetime
  Physics from the Grassmannian},''
  \href{http://dx.doi.org/10.1007/JHEP01(2011)108}{{\em JHEP} {\bf 1101} (2011)
   108},
\href{http://arxiv.org/abs/0912.3249}{{ arXiv:0912.3249 [hep-th]}}.

\bibitem{1312.2007}
N.~Arkani-Hamed and J.~Trnka, ``{The Amplituhedron},''
  \href{https://doi.org/10.1007/JHEP10(2014)030}{{\em JHEP}
  {\bf 10} (2014)  030},
\href{http://arxiv.org/abs/1312.2007}{{ arXiv:1312.2007 [hep-th]}}.

\bibitem{1312.7878}
N.~Arkani-Hamed and J.~Trnka, ``{Into the Amplituhedron},''
  \href{https://doi.org/10.1007/JHEP12(2014)182}{{\em JHEP}
  {\bf 12} (2014)  182},
\href{http://arxiv.org/abs/1312.7878}{{ arXiv:1312.7878 [hep-th]}}.

\bibitem{hep-th/0607160}
J.~M. Drummond, J.~M. Henn, V.~A. Smirnov, E. Sokatchev ``{Magic identities for conformal four-point integrals},''
  \href{http://dx.doi.org/	10.1088/1126-6708/2007/01/064}{{\em JHEP} {\bf 0701}
  (2006)  064},
\href{https://arxiv.org/abs/hep-th/0607160} {{arXiv:0607160 [hep-th]}}.

\bibitem{0707.0243}
J.~M. Drummond, G.~P. Korchemsky, and E.~Sokatchev, ``{Conformal Properties of
  Four-Gluon Planar Amplitudes and Wilson loops},''
  \href{http://dx.doi.org/10.1016/j.nuclphysb.2007.11.041}{{\em Nucl. Phys.}
  {\bf B795} (2008)  385--408},
\href{http://arxiv.org/abs/0707.0243}{{ arXiv:0707.0243 [hep-th]}}.

\bibitem{0807.1095}
J.~Drummond, J.~Henn, G.~Korchemsky, and E.~Sokatchev, ``{Dual Superconformal
  Symmetry of Scattering Amplitudes in $\mathcal{N}\!=\!4$ super Yang-Mills
  Theory},'' \href{http://dx.doi.org/10.1016/j.nuclphysb.2009.11.022}{{\em
  Nucl. Phys.} {\bf B828} (2010)  317--374},
\href{http://arxiv.org/abs/0807.1095}{{ arXiv:0807.1095 [hep-th]}}.

\bibitem{0807.4097}
A.~Brandhuber, P.~Heslop, and G.~Travaglini, ``{A Note on Dual Superconformal
  Symmetry of the $\mathcal{N}\!=\!4$ Super Yang-Mills S-Matrix},''
  \href{http://dx.doi.org/10.1103/PhysRevD.78.125005}{{\em Phys. Rev.} {\bf
  D78} (2008)  125005},
\href{http://arxiv.org/abs/0807.4097}{{ arXiv:0807.4097 [hep-th]}}.

\bibitem{0905.1473}
A.~Hodges, ``{Eliminating Spurious Poles from Gauge-Theoretic Amplitudes},''
  \href{http://dx.doi.org/10.1007/JHEP05(2013)135}{{\em JHEP} {\bf 1305} (2013)
   135},
\href{http://arxiv.org/abs/0905.1473}{{ arXiv:0905.1473 [hep-th]}}.

\bibitem{0912.4912}
N.~Arkani-Hamed, J.~Bourjaily, F.~Cachazo, and J.~Trnka, ``{Unification of
  Residues and Grassmannian Dualities},''
  \href{http://dx.doi.org/10.1007/JHEP01(2011)049}{{\em JHEP} {\bf 1101} (2011)
   049},
\href{http://arxiv.org/abs/0912.4912}{{ arXiv:0912.4912 [hep-th]}}.

\bibitem{1002.4625}
G.~P.~Korchemsky and E.~Sokatchev,
``Superconformal invariants for scattering amplitudes in N=4 SYM theory,''
\href{http://dx.doi.org/10.1016/j.nuclphysb.2010.05.022}{{\em
  Nucl. Phys.} {\bf B839} (2010)  377--419},
\href{http://arxiv.org/abs/1002.4625}{{ arXiv:1002.4625 [hep-th]}}.


\bibitem{1002.4622}
J.~M.~Drummond and L.~Ferro,
``The Yangian origin of the Grassmannian integral,''
 \href{http://dx.doi.org/10.1007/JHEP12(2010)010}{{\em JHEP} {\bf 1012} (2010)
   010},
\href{http://arxiv.org/abs/1002.4622}{{ arXiv:1002.4622 [hep-th]}}.

\bibitem{1212.5605}
N.~Arkani-Hamed, J.~L. Bourjaily, F.~Cachazo, A.~B. Goncharov, A.~Postnikov,
  and J.~Trnka, ``{Scattering Amplitudes and the Positive Grassmannian},''
\href{http://arxiv.org/abs/1212.5605}{{ arXiv:1212.5605 [hep-th]}}.

\bibitem{1006.5703}
A. B. ~Goncharov, M. ~Spradlin, C. ~Vergu, and A. ~Volovich, ``{Classical Polylogarithms for Amplitudes and Wilson Loops},''
  \href{https://doi.org/10.1103/PhysRevLett.105.151605}{{\em Phys. Lett.}
  {\bf 105} (2010)  151605},
\href{http://arxiv.org/abs/1006.5703}{{ arXiv:1006.5703 [hep-th]}}.

\bibitem{0909.0483}
N.~Arkani-Hamed, F.~Cachazo, and C.~Cheung,
``The Grassmannian Origin Of Dual Superconformal Invariance,''
\href{http://dx.doi.org/10.1007/JHEP03(2010)036}{{\em JHEP} {\bf 1003} (2010)
   036},
\href{http://arxiv.org/abs/0909.0483}{{ arXiv:0909.0483 [hep-th]}}.

\bibitem{1701.00453}
B.~Eden, P.~Heslop, and L.~Mason ``{The Correlahedron},''
\href{https://doi.org/10.1007/JHEP09(2017)156}{{\em JHEP}
  {\bf 09} (2017)  156},
  \href{http://arxiv.org/abs/1701.00453}{{ arXiv:1701.00453 [hep-th]}}.

\bibitem{1704.05069}
N.~Arkani-Hamed, H.~Thomas, and J.~Trnka,
``Unwinding the Amplituhedron in Binary,''
\href{http://dx.doi.org/10.1007/JHEP01(2018)016}{{\em JHEP} {\bf 1801} (2018)
   016},
\href{http://arxiv.org/abs/1704.05069}{{ arXiv:1704.05069 [hep-th]}}.

\bibitem{0709.2368}
J.~M. Drummond, J.~Henn, G.~P. Korchemsky, and E.~Sokatchev, ``{On Planar Gluon
  Amplitudes/Wilson Loops Duality},''
  \href{http://dx.doi.org/10.1016/j.nuclphysb.2007.11.007}{{\em Nucl. Phys.}
  {\bf B795} (2008)  52--68},
\href{http://arxiv.org/abs/0709.2368}{{ arXiv:0709.2368 [hep-th]}}.

\bibitem{0902.2987}
J.~M. Drummond, J.~M. Henn, and J.~Plefka, ``{Yangian Symmetry of Scattering
  Amplitudes in $\mathcal{N}\!=\!4$ Super Yang-Mills Theory},''
  \href{http://dx.doi.org/10.1088/1126-6708/2009/05/046}{{\em JHEP} {\bf 05}
  (2009)  046},
\href{http://arxiv.org/abs/0902.2987}{{ arXiv:0902.2987 [hep-th]}}.

\bibitem{0705.0303}
L.~F. Alday, and J.~M. Maldacena, ``{Gluon Scattering Amplitudes at Strong
  Coupling},'' \href{http://dx.doi.org/10.1088/1126-6708/2007/06/064}{{\em
  JHEP} {\bf 06} (2007)  064},
\href{http://arxiv.org/abs/0705.0303}{{ arXiv:0705.0303 [hep-th]}}.

\bibitem{1012.6030}
N.~Arkani-Hamed, J.~L. Bourjaily, F.~Cachazo, A.~Hodges, and J.~Trnka, ``{A Note on Polytopes for Scattering Amplitudes},''
  \href{https://doi.org/10.1007/JHEP04(2012)081}{{\em JHEP}
  {\bf 04} (2012)  081},
\href{http://arxiv.org/abs/1012.6030}{{ arXiv:1012.6030 [hep-th]}}.

\bibitem{1009.2225}
L.~Mason and D.~Skinner, ``{The Complete Planar $S$-Matrix of
  $\mathcal{N}\!=\!4$ SYM as a Wilson Loop in Twistor Space},''
  \href{http://dx.doi.org/10.1007/JHEP12(2010)018}{{\em JHEP} {\bf 12} (2010)
  018},
\href{http://arxiv.org/abs/1009.2225}{{ arXiv:1009.2225 [hep-th]}}.

\bibitem{1010.1167}
S.~Caron-Huot, ``{Notes on the Scattering Amplitude / Wilson Loop Duality},''
  \href{http://dx.doi.org/10.1007/JHEP07(2011)058}{{\em JHEP} {\bf 1107} (2011)
   058},
\href{http://arxiv.org/abs/1010.1167}{{ arXiv:1010.1167 [hep-th]}}.

\bibitem{PHLTA,B214,215}
V.~P. Nair, ``{A Current Algebra for Some Gauge Theory Amplitudes},''
\href{http://dx.doi.org/10.1016/0370-2693(88)91471-2}{{\em Phys. Lett.} {\bf
  B214} (1988)  215--218}.

\bibitem{9702424}
Z.~Bern, J.~Rozowsky, and B.~Yan, ``{Two-Loop Four-Gluon Amplitudes in
  $\mathcal{N}\!=\!4$ Super Yang-Mills},''
  \href{http://dx.doi.org/10.1016/S0370-2693(97)00413-9}{{\em Phys. Lett.} {\bf
  B401} (1997)  273--282},
\href{http://arxiv.org/abs/hep-ph/9702424}{{ arXiv:hep-ph/9702424}}.

\bibitem{0309040}
C.~Anastasiou, Z.~Bern, L.~J. Dixon, and D.~A. Kosower, ``{Planar Amplitudes in
  Maximally Supersymmetric Yang-Mills Theory},''
  \href{http://dx.doi.org/10.1103/PhysRevLett.91.251602}{{\em Phys. Rev. Lett.}
  {\bf 91} (2003)  251602},
\href{http://arxiv.org/abs/hep-th/0309040}{{ arXiv:hep-th/0309040}}.

\bibitem{0505205}
Z.~Bern, L.~J. Dixon, and V.~A. Smirnov, ``{Iteration of Planar Amplitudes in
  Maximally Supersymmetric Yang-Mills Theory at Three Loops and Beyond},''
  \href{http://dx.doi.org/10.1103/PhysRevD.72.085001}{{\em Phys. Rev.} {\bf
  D72} (2005)  085001},
\href{http://arxiv.org/abs/hep-th/0505205}{{ arXiv:hep-th/0505205}}.

\bibitem{0610248}
Z.~Bern, M.~Czakon, L.~J. Dixon, D.~A. Kosower, and V.~A. Smirnov, ``{The
  Four-Loop Planar Amplitude and Cusp Anomalous Dimension in Maximally
  Supersymmetric Yang-Mills Theory},''
  \href{http://dx.doi.org/10.1103/PhysRevD.75.085010}{{\em Phys. Rev.} {\bf
  D75} (2007)  085010},
\href{http://arxiv.org/abs/hep-th/0610248}{{ arXiv:hep-th/0610248 [hep-th]}}.

\bibitem{0705.1864}
Z.~Bern, J.~Carrasco, H.~Johansson, and D.~Kosower, ``{Maximally Supersymmetric
  Planar Yang-Mills Amplitudes at Five Loops},''
  \href{http://dx.doi.org/10.1103/PhysRevD.76.125020}{{\em Phys. Rev.} {\bf
  D76} (2007)  125020},
\href{http://arxiv.org/abs/0705.1864}{{ arXiv:0705.1864 [hep-th]}}.

\bibitem{1210.7709}
Z.~Bern, J.~J. Carrasco, L.~J. Dixon, M.~R. Douglas, M.~von Hippel, and
  H.~Johansson, ``{$D\!=\!5$ Maximally Supersymmetric Yang-Mills Theory
  Diverges at Six Loops},''
  \href{http://dx.doi.org/10.1103/PhysRevD.87.025018}{{\em Phys. Rev.} {\bf
  D87} (2013) no. 2, 025018},
\href{http://arxiv.org/abs/1210.7709}{{ arXiv:1210.7709 [hep-th]}}.

\bibitem{0003203}
M.~Bianchi, S.~Kovacs, G.~Rossi, and Y.~S. Stanev, ``{Anomalous Dimensions in
  $\mathcal{N}\!=\!4$ SYM Theory at Order $g^4$},''
  \href{http://dx.doi.org/10.1016/S0550-3213(00)00312-6}{{\em Nucl. Phys.} {\bf
  B584} (2000)  216--232},
\href{http://arxiv.org/abs/hep-th/0003203}{{ arXiv:hep-th/0003203 [hep-th]}}.

\bibitem{1202.5733}
B.~Eden, P.~Heslop, G.~P. Korchemsky, V.~A. Smirnov, and E.~Sokatchev,
  ``{Five-Loop Konishi in $\mathcal{N}\!=\!4$ SYM},''
  \href{http://dx.doi.org/10.1016/j.nuclphysb.2012.04.015}{{\em Nucl. Phys.}
  {\bf B862} (2012)  123--166},
\href{http://arxiv.org/abs/1202.5733}{{arXiv:1202.5733 [hep-th]}}.

\bibitem{0808.0491}
J.~Drummond, J.~Henn, G.~Korchemsky, and E.~Sokatchev, ``{Generalized unitarity for $\mathcal{N}\!=\!4$ super-amplitudes},'' \href{http://dx.doi.org/10.1016/j.nuclphysb.2012.12.009}{{\em
  Nucl. Phys.} {\bf B869} (2013)  452--492},
\href{http://arxiv.org/abs/0808.0491}{{ arXiv:0808.0491 [hep-th]}}.

\end{thebibliography}
\end{document}